\documentclass[final,3p,sort,compress,times]{elsarticle}

\usepackage{multirow,setspace,times,amssymb,amsmath,graphicx,color,rotating,subfigure,url}

\begin{document}

\begin{frontmatter}

\title{Time series momentum and contrarian effects in the Chinese stock market}

\author[BS,RCE]{Huai-Long Shi}
\author[BS,RCE,SS]{Wei-Xing Zhou\corref{cor1}}
\cortext[cor1]{Corresponding author. Address: 130 Meilong Road, P.O. Box 114, School of Business, East China University of Science and Technology, Shanghai 200237, China, Phone: +86-21-64253634.}
\ead{wxzhou@ecust.edu.cn}%

\address[BS]{Department of Finance, School of Business, East China University of Science and Technology, Shanghai 200237, China}
\address[RCE]{Research Center for Econophysics, East China University of Science and Technology, Shanghai 200237, China}
\address[SS]{Department of Mathematics, School of Science, East China University of Science and Technology, Shanghai 200237, China}

\begin{abstract}
  This paper concentrates on the time series momentum or contrarian effects in the Chinese stock market. We evaluate the performance of the time series momentum strategy applied to major stock indices in mainland China and explore the relation between the performance of time series momentum strategies and some firm-specific characteristics. Our findings indicate that there is a time series momentum effect in the short run and a contrarian effect in the long run in the Chinese stock market. The performances of the time series momentum and contrarian strategies are highly dependent on the look-back and holding periods and firm-specific characteristics.
\end{abstract}


\end{frontmatter}

\section{Introduction}

Market anomalies provide the direct evidence against the Efficient Markets Hypothesis (EMH). Among the anomalies, there are two renowned phenomena, namely, the momentum and contrarian effects, which are wildly investigated. There two versions of momentum or contrarian effects. The earlier and first version is the cross-sectional momentum and contrarian effects \cite{Jegadeesh-Titman-1993-JF,DeBondt-Thaler-1985-JF}. The cross-sectional momentum and contrarian effects are found to be ubiquitous in many asset classes as well as in most of regions around the world. Although the majority of earlier studies effects focus on the U.S. market \cite{DeBondt-Thaler-1985-JF,Jegadeesh-Titman-1993-JF,Jegadeesh-Titman-2001-JF,Gutierrez-Kelley-2008-JF}, a large body of research reported evidence of the momentum (contrarian) effects in other regions or markets, including the UK \cite{Hon-Tonks-2003-JMFM,Galariotis-Holmes-Ma-2007-JMFM}, Japan \cite{Chou-Wei-Chung-2007-JEF}, Australia \cite{Demir-Muthuswamy-Walter-2004-PBFJ}, China \cite{Kang-Liu-Ni-2002-PBFJ,Wang-Chin-2004-PBFJ,Naughton-Truong-Veeraraghavan-2008-PBFJ,Pan-Tang-Xu-2013-PBFJ,Shi-Jiang-Zhou-2015-PLoS1,Shi-Jiang-Zhou-2017-RAPS}, to list a few. Moreover, the momentum and contrarian effects are also found in many other asset classes \cite{NovyMarx-2012-JFE,Asness-Moskowitz-Pedersen-2013-JF}, including funds \cite{Grinblatt-Titman-Wermers-1995-AER,Carhart-1997-JF}, currencies \cite{Kho-1996-JFE,Nitschka-2010-GER,Menkhoff-Sarno-Schmeling-Schrimpf-2012-JFE},
and commodities \cite{Erb-Harvey-2006-FAJ,Miffre-Rallis-2007-JBF}.

In essence, the cross-sectional momentum (contrarian) portfolios are zero-cost arbitrage portfolios that are constructed via buying (selling) winners and selling (buying) losers.
Apart from the cross-sectional momentum (contrarian) effect, a newer version of momentum anomaly were unveiled recently \cite{Moskowitz-Ooi-Pedersen-2012-JFE}, that is, the time series momentum effect (briefly, the TSMOM effect). It describes the predictability of future price trend of a single asset based on its past own performance. Evidence of a trend-following phenomenon has been documented for 58 futures and forward contracts from different asset classes, and it is argued that the TSMOM effect represents one of the most direct tests of the random walk hypothesis and a number of prominent behavioral and rational asset pricing theories \cite{Moskowitz-Ooi-Pedersen-2012-JFE}.

Compared with the cross-sectional momentum (contrarian) effect, studies on the TSMOM effect are relatively fewer, most of which focus on the futures or forwards markets
\cite{Moskowitz-Ooi-Pedersen-2012-JFE,Clare-Seaton-Smith-Thomas-2014-IRFA,Pettersson-2014-SSRN,Kim-Tse-Wald-2016-JFinM}, mainly owing to less constraint of short selling as well as less trading cost. Meanwhile, a growing body of research directs attention on the stock market, particularly on stock indexes \cite{He-Li-2015-JBF,Chakrabarti-2015-BE,DSouza-Srichanachaichok-Wang-Yao-2016-SSRN}.
Most of aforementioned works follow the trading signaling process derived in Ref.~\cite{Moskowitz-Ooi-Pedersen-2012-JFE} to study the TSMOM effect by holding the long (short) positions if the average excess return over past periods is positive (negative). Alternatively, a slightly different signaling method has been proposed with an indicator function of decaying weighted average of past return based on the evidence of the TSMOM effect in the S\&P 500 index \cite{He-Li-2015-JBF}.

Our work contributes to the literature by taking a close look at the time series momentum or contrarian effects in the Chinese stock market. Compared with mature financial markets, the Chinese stock market is relatively young. Some idiosyncratic phenomena characterize the Chinese market, including the less transparent information environment at the market level and the firm level and a larger proportion of irrational individual investors \cite{Zhang-Li-Shen-Teglio-2016-EM}. We wonder whether the time series momentum (contrarian) effect is more pronounced in the Chinese market. In addition, to our knowledge, few studies on the TSMOM effect have paid attention to the Chinese market. Our study aims to provide more empirical evidence from the Chinese market to the literature on the TSMOM effect.

We attempt to testify whether there is the time series momentum or contrarian effects in the Chinese stock market. Specifically, we evaluate the performance of the TSMOM strategies with various look-back and holding periods applied to the Chinese major stock indices. Due to the aforementioned immature market characteristics, we expect the more pronounced anomalies in the Chinese stock market.
We conjecture that the times-series momentum or contrarian effect is related to firm-specific characteristics. As documented in Ref.~\cite{Shi-Jiang-Zhou-2015-PLoS1}, the cross-sectional momentum effects are related to firm-specific characteristics such as market value \cite{Jegadeesh-Titman-1993-JF,Rouwenhorst-1998-JF}, price \cite{Lesmond-Schill-Zhou-2004-JFE}, trading volume \cite{Lee-Swaminathan-2000-JF}, or turnover rate \cite{Naughton-Truong-Veeraraghavan-2008-PBFJ}. Intuitively, the newer version of momentum anomaly TSMOM is also related to some firm-specific characteristics. Hence, we intend to explore the relation of the TSMOM effect and firm-specific factors, including closing price, adjusted price, market capitalization, turnover rate, trading volume, etc.
We also conjecture that the time series momentum or contrarian effect based on various industrial sectors differs. It is argued that individual stock momentum is a function of momentum within industry returns \cite{Moskowitz-Grinblatt-1999-JF}, and the cross-sectional momentum profits are mainly from industry momentum portfolios that are constructed by buying the stocks from past winner industry and selling the stocks from past loser industry. In this vein, we also intend to explore the relation of industry and time series momentum (contrarian) effects via examining the respective performance of the TSMOM strategies based on various industrial sectors.

The remainder of paper is arranged as follows. Section \ref{S1:Method} and Section \ref{S1:Data} describe the methodology and data, respectively. Section \ref{S1:Results} presents the empirical findings. Section \ref{S1:Conclusion} concludes.

\section{Methodology}
\label{S1:Method}

Following the frequently used procedure on studying the cross-sectional momentum (contrarian) effects, we study the TSMOM strategy through constructing $J$-$K$ strategies, where $J$ and $K$ represent respectively the look-back and holding period.
For the robustness of results, two trading signaling methods are employed in our work. Specifically, in light of Ref.~\cite{Moskowitz-Ooi-Pedersen-2012-JFE}, the positive (negative) excess return for equity $i$ over the past $J$ months implies buying (selling) the equity $i$ in the future. Accordingly, the signaling process can be described as follows,
\begin{equation}
  TS_{t}^{i}({\mathrm{MOP}})=\mathrm{sign}\left\{\frac{1}{J}\left[\left(r_{t-1}^{i}-r_{f,t-1}\right)+ \cdots +\left(r_{t-J}^{i}-r_{f,t-J}\right)\right]\right\},
  \label{Eq:Signal:v1}
\end{equation}
where $r_t=\ln(P_t^{i})-\ln(P_{t-1}^{i})$ and $r_{f,t}$ is the 1 month Treasury bill rate of the $t$-th month.
An alternative method is derived from Ref.~\cite{He-Li-2015-JBF}, which is implied by an indicator function of decaying weighted average of past return over the look-back period, as described in the following,
\begin{equation}
  TS_{t}^{i}({\mathrm{HL}})=\mathrm{sign}\left\{\frac{1}{J}\left[Jr_{t-1}^{i}+(J-1)r_{t-2}^{i}+ \cdots +r_{t-J}^{i}\right]\right\}.
  \label{Eq:Signal:v2}
\end{equation}
The trading signaling method ${\mathrm{HL}}$ can be equivalently viewed as the indicator function of moving average process of the return time series $r_t^i$ .

Accordingly, the average monthly excess return to the TSMOM strategy could be depicted calculated as follows,
\begin{equation}
\centering
\left[\frac{1}{K}\sum_{m=1}^{K}TS_{t-m}^{i}\right]\times (r_{t}^{i}-r_{f,t}).
\label{Eq:TSMOM}
\end{equation}
In other words, we can earn either the return of the TSMOM strategies or the return that is the risk-free interest rate. In the latter case, there is zero excess returns \cite{Clare-Seaton-Smith-Thomas-2014-IRFA}.

We then obtain the $t$-statistics of the return series, adjusted for heteroscedasticity and autocorrelation \cite{Newey-West-1987-Em}, to verify whether it is the time series momentum (TSMOM) effect or the time series contrarian (TSCON) effect. It would be the significant TSMOM (TSCON) effect if the $t$-value is positive (negative) with statistical significance. All the strategies would be divided into two groups, that is, the strategies that incur positive returns of statistical significance and the strategies that result in negative returns of statistical significance.

As argued in Ref.~\cite{Gary-2013-SSEP}, stock indices are less subject to liquidity and microstructure issues. Since our work mainly concentrate on the stock indices, we do not skip one month between the look-back and holding periods, which is often used in the literature about the cross-sectional momentum (contrarian) effects for the correction of measurement errors \cite{Jegadeesh-Titman-1993-JF,Jegadeesh-Titman-2001-JF,Kang-Liu-Ni-2002-PBFJ,Shi-Jiang-Zhou-2015-PLoS1}.
According to Ref.~\cite{Kim-Tse-Wald-2016-JFinM}, the profitability of the TSMOM effect in Ref.~\cite{Moskowitz-Ooi-Pedersen-2012-JFE} is attributed to volatility scaling or risk parity approach, and the risk parity approach gives rise to the superior performance of the TSMOM strategy. They also found that the TSMOM strategy and the buy \& hold strategy share the similar performance, no matter whether the risk parity approach is adopted. Hence, we only concentrate on the pure TSMOM (TSCON) effect without volatility scaling.

\section{Data sets}
\label{S1:Data}

We retrieve the data from the RESSET database (http://www.resset.cn), which provides domestic economic and financial data for many colleges and academic institutions in China. Our data sets consist of three stock indexes, including Shanghai Stock Exchange Composite Index (SHCI), Shenzhen Stock Exchange Component Index (SZCI), and Shanghai Shenzhen 300 Index (CSI 300), and all A-share individual stocks in Shanghai Stock Exchange (SHSE) and Shenzhen Stock Exchange (SZSE). The A-share individual stock data consists of the monthly closing price, adjusted price (for dividend and splitting), market value (based on both outstanding share and all share), turnover rate and trading volume, covering the period from 1991 to 2015.

\section{Empirical findings}
\label{S1:Results}

This section presents the results with various look-back ($J$) and holding horizon ($K$) over the whole sample period, and $J,K \in \{1,3,6,9,12,24,36,48,60\}$ month(s). Accordingly, we consider a total of 81 ($9 \times 9$) TSMOM strategies. We examine the respective performances of the TSMOM strategies applied to SHCI, SZCI, and CSI 300 as well as the stock group ``index'' in terms of some firm-specific characteristics. We also evaluate the performance of the TSMOM strategies applied to stock groups based on industrial sectors.

\subsection{The performance of major stock indices}

At first, we obtain some basic results for major stock indices of the Chinese stock markets. The three panels of Table \ref{TB:ChinaIndex:TSMOM} report the results for SHCI, SZCI, and CSI 300, respectively. We observe both the TSMOM and the TSCON effects for the three stock indices, irrespective of the trading signaling methods HL and MOP. There are more TSCON strategies that are statistically significant, which suggests that the TSCON effect prevails over the TSMOM strategies. The TSCON effect exists when the look-back or holding period in the long-term, while the TSMOM effect occurs when the look-back and holding horizons are in short term, especially with $J$ and $K$ ranging from $1$-month to $12$-month. As with the viable explanation of cross-sectional momentum (contrarian) effect, it is more likely that the under-reaction to the news gives rise to the short-term TSMOM effect, and the overreaction leads to the long-term TSCON effect. It is also noticeable that the return magnitude based on short-term look-back and holding periods (especially when $J,K<12$), are much larger than the rest. In other words, significant TSMOM strategies perform better than significant TSCON strategies. We also notice that the significant TSMOM/TSCON effects are more widespread in the cases of SZCI and CSI 300. This may be attributed to the fact that the average capitalization is smaller for the stocks on the Shenzhen Stock Exchange.

\setlength\tabcolsep{0.7pt}
\begin{table}[!ht]
\centering
  \caption{(Color online) This table reports the annualized monthly excess returns to TSMOM strategy based on the HL and MOP methods, applied to three major stock indices (SHCI, SZCI and CSI 300). The look-back and holding periods are $J$ and $K$, respectively. Following Ref.~\cite{Newey-West-1987-Em}, the $t$-statistics are adjusted for heteroscedasticity and autocorrelation. The superscripts * and ** denote the significance at 5\% and 1\% levels, respectively.}
  \medskip
\footnotesize
   \begin{tabular}{lccccccccccccccccccccc}
   \hline
   && \multicolumn{9}{c}{HL}  &&& \multicolumn{9}{c}{MOP} \\
   \cline{3-11} \cline{14-22}
    $J$ && $K=1$ &3 & 6 & 9 & 12 & 24 & 36 & 48 & 60 &&& 1 & 3 & 6 & 9 & 12 & 24 & 36 & 48 & 60\\
   \hline
   \vspace{-3mm}\\
   \multicolumn{22}{l}{\textit{Panel A: SHCI}} \\
   $1$ && {\color{red} 0.16$^{~~}$} & {\color{red} 0.13$^{~~}$} & {\color{red} 0.07$^{~~}$} & {\color{red} 0.08$^{~~}$} & {\color{red} 0.07$^{~~}$} & {\color{red} 0.03$^{~~}$} & {\color{red} 0.02$^{~~}$} & {\color{red} 0.03$^{~~}$} & {\color{red} 0.02$^{~~}$}&&& {\color{red} 0.17$^{~~}$} & {\color{red} 0.14$^{~~}$} & {\color{red} 0.07$^{~~}$} & {\color{red} 0.08$^{~~}$} & {\color{red} 0.07$^{~~}$} & {\color{red} 0.03$^{~~}$} & {\color{red} 0.02$^{~~}$} & {\color{red} 0.03$^{~~}$} & {\color{red} 0.03$^{~~}$} \\
   $3$ && {\color{red} 0.17$^{~~}$} & {\color{red} 0.13$^{~~}$} & {\color{red} 0.07$^{~~}$} & {\color{red} 0.09$^{~~}$} & {\color{red} 0.06$^{~~}$} & {\color{blue}-0.01$^{~~}$} & {\color{blue}-0.01$^{~~}$} & {\color{red} 0.00$^{~~}$} & {\color{red} 0.00$^{~~}$}&&& {\color{red} \bf  0.21$^{*~}$} & {\color{red} 0.12$^{~~}$} & {\color{red} 0.07$^{~~}$} & {\color{red} 0.08$^{~~}$} & {\color{red} 0.05$^{~~}$} & {\color{blue}-0.01$^{~~}$} & {\color{blue}-0.01$^{~~}$} & {\color{blue}-0.00$^{~~}$} & {\color{blue}-0.01$^{~~}$} \\
   $6$ && {\color{red} 0.10$^{~~}$} & {\color{red} 0.06$^{~~}$} & {\color{red} 0.06$^{~~}$} & {\color{red} 0.09$^{~~}$} & {\color{red} 0.07$^{~~}$} & {\color{red} 0.04$^{~~}$} & {\color{red} 0.04$^{~~}$} & {\color{red} 0.05$^{~~}$} & {\color{red} 0.04$^{~~}$}&&& {\color{red} 0.07$^{~~}$} & {\color{red} 0.07$^{~~}$} & {\color{red} 0.09$^{~~}$} & {\color{red} 0.10$^{~~}$} & {\color{red} 0.06$^{~~}$} & {\color{red} 0.03$^{~~}$} & {\color{red} 0.03$^{~~}$} & {\color{red} 0.04$^{~~}$} & {\color{red} 0.03$^{~~}$} \\
   $9$ && {\color{red} 0.09$^{~~}$} & {\color{red} 0.08$^{~~}$} & {\color{red} 0.07$^{~~}$} & {\color{red} 0.08$^{~~}$} & {\color{red} 0.04$^{~~}$} & {\color{red} 0.02$^{~~}$} & {\color{red} 0.02$^{~~}$} & {\color{red} 0.03$^{~~}$} & {\color{red} 0.02$^{~~}$}&&& {\color{red} 0.10$^{~~}$} & {\color{red} 0.09$^{~~}$} & {\color{red} 0.07$^{~~}$} & {\color{red} 0.04$^{~~}$} & {\color{red} 0.03$^{~~}$} & {\color{red} 0.01$^{~~}$} & {\color{red} 0.02$^{~~}$} & {\color{red} 0.02$^{~~}$} & {\color{red} 0.01$^{~~}$} \\
   $12$ && {\color{red} 0.09$^{~~}$} & {\color{red} 0.09$^{~~}$} & {\color{red} 0.06$^{~~}$} & {\color{red} 0.06$^{~~}$} & {\color{red} 0.03$^{~~}$} & {\color{red} 0.00$^{~~}$} & {\color{red} 0.01$^{~~}$} & {\color{red} 0.01$^{~~}$} & {\color{red} 0.00$^{~~}$}&&& {\color{red} 0.11$^{~~}$} & {\color{red} 0.10$^{~~}$} & {\color{red} 0.05$^{~~}$} & {\color{red} 0.04$^{~~}$} & {\color{red} 0.03$^{~~}$} & {\color{blue}-0.02$^{~~}$} & {\color{blue}-0.00$^{~~}$} & {\color{blue}-0.00$^{~~}$} & {\color{blue}-0.01$^{~~}$} \\
   $24$ && {\color{red} 0.04$^{~~}$} & {\color{red} 0.02$^{~~}$} & {\color{blue}-0.02$^{~~}$} & {\color{blue}-0.02$^{~~}$} & {\color{blue}-0.03$^{~~}$} & {\color{blue}-0.06$^{~~}$} & {\color{blue}-0.05$^{~~}$} & {\color{blue}-0.04$^{~~}$} & {\color{blue}-0.05$^{~~}$}&&& {\color{blue}-0.05$^{~~}$} & {\color{blue}-0.06$^{~~}$} & {\color{blue}-0.04$^{~~}$} & {\color{blue}-0.05$^{~~}$} & {\color{blue}-0.06$^{~~}$} & {\color{blue}-0.08$^{~~}$} & {\color{blue}-0.05$^{~~}$} & {\color{blue}-0.05$^{~~}$} & {\color{blue}-0.06$^{~~}$} \\
   $36$ && {\color{red} 0.00$^{~~}$} & {\color{blue}-0.04$^{~~}$} & {\color{blue}-0.04$^{~~}$} & {\color{blue}-0.04$^{~~}$} & {\color{blue}-0.06$^{~~}$} & {\color{blue}-0.08$^{~~}$} & {\color{blue}-0.06$^{~~}$} & {\color{blue} \bf -0.07$^{*~}$} & {\color{blue} \bf -0.08$^{*~}$}&&& {\color{red} 0.03$^{~~}$} & {\color{red} 0.01$^{~~}$} & {\color{blue}-0.01$^{~~}$} & {\color{blue}-0.04$^{~~}$} & {\color{blue}-0.04$^{~~}$} & {\color{blue}-0.06$^{~~}$} & {\color{blue}-0.05$^{~~}$} & {\color{blue}-0.07$^{~~}$} & {\color{blue}-0.06$^{~~}$} \\
   $48$ && {\color{red} 0.01$^{~~}$} & {\color{blue}-0.00$^{~~}$} & {\color{blue}-0.01$^{~~}$} & {\color{blue}-0.02$^{~~}$} & {\color{blue}-0.02$^{~~}$} & {\color{blue}-0.04$^{~~}$} & {\color{blue}-0.04$^{~~}$} & {\color{blue}-0.05$^{~~}$} & {\color{blue}-0.04$^{~~}$}&&& {\color{blue}-0.03$^{~~}$} & {\color{blue}-0.03$^{~~}$} & {\color{blue}-0.03$^{~~}$} & {\color{blue}-0.03$^{~~}$} & {\color{blue}-0.04$^{~~}$} & {\color{blue}-0.05$^{~~}$} & {\color{blue}-0.05$^{~~}$} & {\color{blue}-0.04$^{~~}$} & {\color{blue}-0.02$^{~~}$} \\
   $60$ && {\color{red} 0.04$^{~~}$} & {\color{red} 0.02$^{~~}$} & {\color{red} 0.01$^{~~}$} & {\color{blue}-0.00$^{~~}$} & {\color{blue}-0.01$^{~~}$} & {\color{blue}-0.03$^{~~}$} & {\color{blue}-0.02$^{~~}$} & {\color{blue}-0.02$^{~~}$} & {\color{blue}-0.01$^{~~}$}&&& {\color{blue}-0.02$^{~~}$} & {\color{blue}-0.05$^{~~}$} & {\color{blue}-0.04$^{~~}$} & {\color{blue}-0.05$^{~~}$} & {\color{blue}-0.06$^{~~}$} & {\color{blue}-0.06$^{~~}$} & {\color{blue}-0.04$^{~~}$} & {\color{blue}-0.01$^{~~}$} & {\color{red} 0.00$^{~~}$} \\
   \hline
   \multicolumn{22}{l}{\textit{Panel B: SZCI}} \\
   $1$ && {\color{red} \bf  0.20$^{*~}$} & {\color{red} \bf  0.14$^{*~}$} & {\color{red} 0.08$^{~~}$} & {\color{red} 0.07$^{~~}$} & {\color{red} 0.05$^{~~}$} & {\color{blue}-0.01$^{~~}$} & {\color{blue}-0.01$^{~~}$} & {\color{blue}-0.00$^{~~}$} & {\color{blue}-0.01$^{~~}$}&&& {\color{red} \bf  0.18$^{**}$} & {\color{red} \bf  0.13$^{*~}$} & {\color{red} 0.08$^{~~}$} & {\color{red} 0.07$^{~~}$} & {\color{red} 0.05$^{~~}$} & {\color{blue}-0.02$^{~~}$} & {\color{blue}-0.02$^{~~}$} & {\color{blue}-0.01$^{~~}$} & {\color{blue}-0.01$^{~~}$} \\
   $3$ && {\color{red} \bf  0.25$^{**}$} & {\color{red} 0.14$^{~~}$} & {\color{red} 0.08$^{~~}$} & {\color{red} 0.07$^{~~}$} & {\color{red} 0.04$^{~~}$} & {\color{blue}-0.03$^{~~}$} & {\color{blue}-0.03$^{~~}$} & {\color{blue}-0.02$^{~~}$} & {\color{blue}-0.02$^{~~}$}&&& {\color{red} \bf  0.23$^{*~}$} & {\color{red} 0.11$^{~~}$} & {\color{red} 0.07$^{~~}$} & {\color{red} 0.06$^{~~}$} & {\color{red} 0.02$^{~~}$} & {\color{blue}-0.03$^{~~}$} & {\color{blue}-0.03$^{~~}$} & {\color{blue}-0.02$^{~~}$} & {\color{blue}-0.02$^{~~}$} \\
   $6$ && {\color{red} \bf  0.23$^{*~}$} & {\color{red} 0.10$^{~~}$} & {\color{red} 0.09$^{~~}$} & {\color{red} 0.10$^{~~}$} & {\color{red} 0.06$^{~~}$} & {\color{blue}-0.01$^{~~}$} & {\color{blue}-0.01$^{~~}$} & {\color{red} 0.02$^{~~}$} & {\color{red} 0.01$^{~~}$}&&& {\color{red} 0.10$^{~~}$} & {\color{red} 0.08$^{~~}$} & {\color{red} 0.09$^{~~}$} & {\color{red} 0.06$^{~~}$} & {\color{red} 0.02$^{~~}$} & {\color{blue}-0.04$^{~~}$} & {\color{blue}-0.04$^{~~}$} & {\color{blue}-0.01$^{~~}$} & {\color{blue}-0.02$^{~~}$} \\
   $9$ && {\color{red} 0.15$^{~~}$} & {\color{red} 0.10$^{~~}$} & {\color{red} 0.09$^{~~}$} & {\color{red} 0.08$^{~~}$} & {\color{red} 0.03$^{~~}$} & {\color{blue}-0.05$^{~~}$} & {\color{blue}-0.03$^{~~}$} & {\color{blue}-0.00$^{~~}$} & {\color{blue}-0.01$^{~~}$}&&& {\color{red} 0.15$^{~~}$} & {\color{red} 0.13$^{~~}$} & {\color{red} 0.09$^{~~}$} & {\color{red} 0.03$^{~~}$} & {\color{red} 0.01$^{~~}$} & {\color{blue}-0.06$^{~~}$} & {\color{blue}-0.04$^{~~}$} & {\color{blue}-0.00$^{~~}$} & {\color{blue}-0.02$^{~~}$} \\
   $12$ && {\color{red} 0.13$^{~~}$} & {\color{red} 0.11$^{~~}$} & {\color{red} 0.06$^{~~}$} & {\color{red} 0.03$^{~~}$} & {\color{blue}-0.02$^{~~}$} & {\color{blue} \bf -0.09$^{*~}$} & {\color{blue}-0.07$^{~~}$} & {\color{blue}-0.03$^{~~}$} & {\color{blue}-0.04$^{~~}$}&&& {\color{red} 0.15$^{~~}$} & {\color{red} 0.08$^{~~}$} & {\color{red} 0.02$^{~~}$} & {\color{blue}-0.01$^{~~}$} & {\color{blue}-0.03$^{~~}$} & {\color{blue} \bf -0.10$^{*~}$} & {\color{blue}-0.07$^{~~}$} & {\color{blue}-0.05$^{~~}$} & {\color{blue} \bf -0.06$^{*~}$} \\
   $24$ && {\color{red} 0.04$^{~~}$} & {\color{red} 0.00$^{~~}$} & {\color{blue}-0.05$^{~~}$} & {\color{blue}-0.08$^{~~}$} & {\color{blue}-0.09$^{~~}$} & {\color{blue} \bf -0.13$^{**}$} & {\color{blue} \bf -0.08$^{*~}$} & {\color{blue} \bf -0.07$^{*~}$} & {\color{blue} \bf -0.09$^{**}$}&&& {\color{blue}-0.02$^{~~}$} & {\color{blue}-0.06$^{~~}$} & {\color{blue}-0.08$^{~~}$} & {\color{blue}-0.10$^{~~}$} & {\color{blue}-0.11$^{~~}$} & {\color{blue} \bf -0.13$^{*~}$} & {\color{blue}-0.06$^{~~}$} & {\color{blue} \bf -0.07$^{*~}$} & {\color{blue} \bf -0.09$^{**}$} \\
   $36$ && {\color{red} 0.01$^{~~}$} & {\color{blue}-0.04$^{~~}$} & {\color{blue}-0.06$^{~~}$} & {\color{blue}-0.06$^{~~}$} & {\color{blue}-0.06$^{~~}$} & {\color{blue}-0.09$^{~~}$} & {\color{blue}-0.06$^{~~}$} & {\color{blue}-0.07$^{~~}$} & {\color{blue} \bf -0.08$^{*~}$}&&& {\color{blue}-0.03$^{~~}$} & {\color{blue}-0.04$^{~~}$} & {\color{blue}-0.05$^{~~}$} & {\color{blue}-0.05$^{~~}$} & {\color{blue}-0.06$^{~~}$} & {\color{blue}-0.07$^{~~}$} & {\color{blue}-0.05$^{~~}$} & {\color{blue} \bf -0.07$^{*~}$} & {\color{blue} \bf -0.07$^{*~}$} \\
   $48$ && {\color{red} 0.05$^{~~}$} & {\color{red} 0.03$^{~~}$} & {\color{blue}-0.01$^{~~}$} & {\color{blue}-0.02$^{~~}$} & {\color{blue}-0.03$^{~~}$} & {\color{blue}-0.09$^{~~}$} & {\color{blue}-0.08$^{~~}$} & {\color{blue} \bf -0.09$^{*~}$} & {\color{blue} \bf -0.10$^{*~}$}&&& {\color{red} 0.02$^{~~}$} & {\color{red} 0.01$^{~~}$} & {\color{blue}-0.01$^{~~}$} & {\color{blue}-0.01$^{~~}$} & {\color{blue}-0.01$^{~~}$} & {\color{blue}-0.04$^{~~}$} & {\color{blue}-0.05$^{~~}$} & {\color{blue}-0.06$^{~~}$} & {\color{blue} \bf -0.06$^{*~}$} \\
   $60$ && {\color{red} 0.09$^{~~}$} & {\color{red} 0.07$^{~~}$} & {\color{red} 0.04$^{~~}$} & {\color{red} 0.04$^{~~}$} & {\color{red} 0.03$^{~~}$} & {\color{blue}-0.02$^{~~}$} & {\color{blue}-0.02$^{~~}$} & {\color{blue}-0.02$^{~~}$} & {\color{blue}-0.03$^{~~}$}&&& {\color{red} 0.02$^{~~}$} & {\color{blue}-0.01$^{~~}$} & {\color{blue}-0.01$^{~~}$} & {\color{blue}-0.02$^{~~}$} & {\color{blue}-0.03$^{~~}$} & {\color{blue}-0.05$^{~~}$} & {\color{blue}-0.03$^{~~}$} & {\color{blue}-0.03$^{~~}$} & {\color{blue}-0.03$^{~~}$} \\
   \hline
   \multicolumn{22}{l}{\textit{Panel C: CSI 300}} \\
   $1$ && {\color{red} 0.12$^{~~}$} & {\color{red} 0.09$^{~~}$} & {\color{red} 0.05$^{~~}$} & {\color{red} 0.06$^{~~}$} & {\color{red} 0.05$^{~~}$} & {\color{blue}-0.02$^{~~}$} & {\color{blue}-0.01$^{~~}$} & {\color{blue}-0.01$^{~~}$} & {\color{blue}-0.01$^{~~}$}&&& {\color{red} 0.13$^{~~}$} & {\color{red} 0.10$^{~~}$} & {\color{red} 0.06$^{~~}$} & {\color{red} 0.06$^{~~}$} & {\color{red} 0.05$^{~~}$} & {\color{blue}-0.01$^{~~}$} & {\color{blue}-0.01$^{~~}$} & {\color{blue}-0.01$^{~~}$} & {\color{blue}-0.02$^{~~}$} \\
   $3$ && {\color{red} 0.21$^{~~}$} & {\color{red} 0.18$^{~~}$} & {\color{red} 0.08$^{~~}$} & {\color{red} 0.06$^{~~}$} & {\color{red} 0.01$^{~~}$} & {\color{blue}-0.03$^{~~}$} & {\color{blue}-0.04$^{~~}$} & {\color{blue}-0.02$^{~~}$} & {\color{blue} \bf -0.03$^{**}$}&&& {\color{red} \bf  0.25$^{*~}$} & {\color{red} \bf  0.20$^{*~}$} & {\color{red} 0.10$^{~~}$} & {\color{red} 0.09$^{~~}$} & {\color{red} 0.03$^{~~}$} & {\color{blue}-0.03$^{~~}$} & {\color{blue}-0.03$^{~~}$} & {\color{blue}-0.02$^{~~}$} & {\color{blue} \bf -0.03$^{*~}$} \\
   $6$ && {\color{red} \bf  0.23$^{*~}$} & {\color{red} 0.13$^{~~}$} & {\color{red} 0.08$^{~~}$} & {\color{red} 0.07$^{~~}$} & {\color{red} 0.02$^{~~}$} & {\color{blue}-0.03$^{~~}$} & {\color{blue}-0.04$^{~~}$} & {\color{blue}-0.03$^{~~}$} & {\color{blue} \bf -0.03$^{**}$}&&& {\color{red} 0.17$^{~~}$} & {\color{red} 0.10$^{~~}$} & {\color{red} 0.08$^{~~}$} & {\color{red} 0.04$^{~~}$} & {\color{blue}-0.01$^{~~}$} & {\color{blue}-0.04$^{~~}$} & {\color{blue}-0.05$^{~~}$} & {\color{blue}-0.03$^{~~}$} & {\color{blue} \bf -0.04$^{**}$} \\
   $9$ && {\color{red} 0.19$^{~~}$} & {\color{red} 0.12$^{~~}$} & {\color{red} 0.07$^{~~}$} & {\color{red} 0.05$^{~~}$} & {\color{blue}-0.00$^{~~}$} & {\color{blue}-0.05$^{~~}$} & {\color{blue}-0.04$^{~~}$} & {\color{blue}-0.03$^{~~}$} & {\color{blue} \bf -0.04$^{*~}$}&&& {\color{red} 0.12$^{~~}$} & {\color{red} 0.08$^{~~}$} & {\color{red} 0.04$^{~~}$} & {\color{blue}-0.01$^{~~}$} & {\color{blue}-0.02$^{~~}$} & {\color{blue}-0.06$^{~~}$} & {\color{blue}-0.05$^{~~}$} & {\color{blue}-0.04$^{~~}$} & {\color{blue}-0.05$^{~~}$} \\
   $12$ && {\color{red} 0.16$^{~~}$} & {\color{red} 0.11$^{~~}$} & {\color{red} 0.05$^{~~}$} & {\color{red} 0.01$^{~~}$} & {\color{blue}-0.02$^{~~}$} & {\color{blue}-0.05$^{~~}$} & {\color{blue}-0.05$^{~~}$} & {\color{blue}-0.04$^{~~}$} & {\color{blue} \bf -0.05$^{*~}$}&&& {\color{red} 0.14$^{~~}$} & {\color{red} 0.04$^{~~}$} & {\color{blue}-0.01$^{~~}$} & {\color{blue}-0.03$^{~~}$} & {\color{blue}-0.03$^{~~}$} & {\color{blue}-0.08$^{~~}$} & {\color{blue}-0.06$^{~~}$} & {\color{blue}-0.05$^{~~}$} & {\color{blue}-0.05$^{~~}$} \\
   $24$ && {\color{red} 0.02$^{~~}$} & {\color{red} 0.05$^{~~}$} & {\color{blue}-0.00$^{~~}$} & {\color{blue}-0.02$^{~~}$} & {\color{blue}-0.05$^{~~}$} & {\color{blue}-0.09$^{~~}$} & {\color{blue}-0.07$^{~~}$} & {\color{blue} \bf -0.06$^{*~}$} & {\color{blue} \bf -0.05$^{*~}$}&&& {\color{red} 0.03$^{~~}$} & {\color{blue}-0.01$^{~~}$} & {\color{blue}-0.02$^{~~}$} & {\color{blue}-0.05$^{~~}$} & {\color{blue}-0.06$^{~~}$} & {\color{blue}-0.10$^{~~}$} & {\color{blue}-0.07$^{~~}$} & {\color{blue}-0.06$^{~~}$} & {\color{blue} \bf -0.07$^{*~}$} \\
   $36$ && {\color{red} 0.02$^{~~}$} & {\color{blue}-0.05$^{~~}$} & {\color{blue}-0.06$^{~~}$} & {\color{blue}-0.07$^{~~}$} & {\color{blue}-0.07$^{~~}$} & {\color{blue}-0.11$^{~~}$} & {\color{blue}-0.08$^{~~}$} & {\color{blue} \bf -0.07$^{*~}$} & {\color{blue} \bf -0.08$^{*~}$}&&& {\color{blue}-0.01$^{~~}$} & {\color{blue}-0.00$^{~~}$} & {\color{blue}-0.03$^{~~}$} & {\color{blue}-0.05$^{~~}$} & {\color{blue}-0.06$^{~~}$} & {\color{blue}-0.11$^{~~}$} & {\color{blue}-0.08$^{~~}$} & {\color{blue} \bf -0.09$^{*~}$} & {\color{blue} \bf -0.09$^{*~}$} \\
   $48$ && {\color{red} 0.03$^{~~}$} & {\color{blue}-0.01$^{~~}$} & {\color{blue}-0.02$^{~~}$} & {\color{blue}-0.03$^{~~}$} & {\color{blue}-0.03$^{~~}$} & {\color{blue}-0.04$^{~~}$} & {\color{blue}-0.03$^{~~}$} & {\color{blue}-0.04$^{~~}$} & {\color{blue}-0.03$^{~~}$}&&& {\color{red} 0.04$^{~~}$} & {\color{blue}-0.01$^{~~}$} & {\color{blue}-0.05$^{~~}$} & {\color{blue}-0.06$^{~~}$} & {\color{blue}-0.08$^{~~}$} & {\color{blue}-0.08$^{~~}$} & {\color{blue}-0.08$^{~~}$} & {\color{blue}-0.08$^{~~}$} & {\color{blue}-0.07$^{~~}$} \\
   $60$ && {\color{red} 0.02$^{~~}$} & {\color{blue}-0.01$^{~~}$} & {\color{blue}-0.03$^{~~}$} & {\color{blue}-0.04$^{~~}$} & {\color{blue}-0.03$^{~~}$} & {\color{blue}-0.04$^{~~}$} & {\color{blue}-0.04$^{~~}$} & {\color{blue}-0.03$^{~~}$} & {\color{blue}-0.02$^{~~}$}&&& {\color{blue}-0.09$^{~~}$} & {\color{blue}-0.09$^{~~}$} & {\color{blue}-0.05$^{~~}$} & {\color{blue}-0.04$^{~~}$} & {\color{blue}-0.04$^{~~}$} & {\color{blue}-0.03$^{~~}$} & {\color{blue}-0.04$^{~~}$} & {\color{blue}-0.03$^{~~}$} & {\color{blue}-0.02$^{~~}$} \\
   \hline
   \end{tabular}
   \label{TB:ChinaIndex:TSMOM}
\end{table}

Besides, our findings also have some intriguing implications. As documented by a body of previous literature, the existence of market anomalies such as cross-sectional and time series momentum (contrarian) effect provides the direct evidence of the failure of the random walk theory and the failure of the weak-form market efficiency theory as well \cite{Fama-1970-JF,Fama-1991-JF}. These anomalies are able to measure the market efficiency. Nevertheless, it should be noted that, in comparison to the results in Ref.~\cite{He-Li-2015-JBF} that more than 50\% TSMOM strategies applied to the S\&P 500 Index have positive excess returns with statistical significance from 1988 to 2012, there is a weak TSMOM effect in the Chinese stock market. In other words, the level of market efficiency is not revealed properly through investigating the TSMOM effects over the whole sample period. A feasible explanation may be related to the data frequency. As is well known, the Chinese stock market is characterized by long-term bears and short-term bulls, while the US market acts the opposite. Thus, as with cross-sectional momentum effect, the trend following strategy like the TSMOM is more likely to be significant with higher data frequency such as daily and weekly. On the other hand, an alternative explanation may be more reasonable, namely, the ``misspecified'' sample period. According to the Adaptive Markets Hypothesis (AMH) \cite{Lo-2004-JPM,Lo-2005-JIC}, the level of market efficiency fluctuates over time but self-improving and the performance of trading strategy and arbitrage opportunity is time-varying as well. Therefore, the finding that the TSMOM effect is stronger in the US market than in the Chinese market such that there is higher (lower) level of market efficiency in the Chinese (US) market merely holds for a certain fixed time period and could not be viewed as the reflection of the real level of market efficiency, which is in fact rising and falling over time.

\subsection{The performance of stock groups based on some firm-specific characteristics}

Intuitively, time series momentum effect ought to be related to similar firm-specific characteristics as the cross-sectional momentum effect. We evaluate the performance of the TSMOM strategies applied to stock groups formed according to several firm-specific characteristics that have been investigated in the previous literature on the cross-sectional counterparts. We consider the factors including closing price, adjusted price (for dividend and split), market value (based on all shares and outstanding shares), turnover rate, and trading volume. Specifically, at the beginning of each month within the sample period, the A-share individual stocks are sorted into quintile groups according to their respective latest factor value. The TSMOM strategy is then applied to the time series of equally-weighted average monthly return of each stock group, from $G1$ (with the highest value) to $G5$ (with the lowest value). That is, all stocks in the same group would be buying or selling at the same time and these stock groups could be viewed as the factor-specific stock indices. The results for the closing price are presented in Table \ref{TB:ChinaGroupIndex:Closing Price} as an example. There are excess returns equaling 0.00, which are not significant. Hence, we do not present the returns multiplied by 100 as in Ref.~\cite{Wang-Xie-Jiang-Stanley-2016-IREF}.

\setlength\tabcolsep{0.7pt}
\begin{table}[!ht]
\centering
  \caption{(Color online) This table reports the annualized monthly excess returns to TSMOM strategy applied to the stock group in terms of the \textit{Closing Price} factor. At each month, the individual stocks are sorted into quintile groups according to their latest value of \textit{Closing Price}. G1 and G5 correspond to the highest and lowest value, respectively.}
  \medskip
\footnotesize
   \begin{tabular}{lccccccccccccccccccccc}
   \hline
   && \multicolumn{9}{c}{$HL$}  &&& \multicolumn{9}{c}{$MOP$} \\
   \cline{3-11} \cline{14-22}
    $J$ & Group & $K=1$ &3 & 6 & 9 & 12 & 24 & 36 & 48 & 60 &&& 1 & 3 & 6 & 9 & 12 & 24 & 36 & 48 & 60\\
   \hline
   \vspace{-3mm}\\
   $1$ &$G1$& {\color{red} \bf  0.19$^{*~}$} & {\color{red} 0.12$^{~~}$} & {\color{red} 0.10$^{~~}$} & {\color{red} \bf  0.11$^{*~}$} & {\color{red} 0.10$^{~~}$} & {\color{red} 0.06$^{~~}$} & {\color{red} 0.05$^{~~}$} & {\color{red} 0.05$^{~~}$} & {\color{red} 0.04$^{~~}$}&&& {\color{red} \bf  0.23$^{*~}$} & {\color{red} \bf  0.16$^{*~}$} & {\color{red} \bf  0.12$^{*~}$} & {\color{red} \bf  0.13$^{*~}$} & {\color{red} \bf  0.11$^{*~}$} & {\color{red} 0.07$^{~~}$} & {\color{red} 0.06$^{~~}$} & {\color{red} 0.05$^{~~}$} & {\color{red} 0.04$^{~~}$} \\
   &$G2$& {\color{red} 0.18$^{~~}$} & {\color{red} 0.14$^{~~}$} & {\color{red} 0.08$^{~~}$} & {\color{red} 0.09$^{~~}$} & {\color{red} 0.09$^{~~}$} & {\color{red} 0.03$^{~~}$} & {\color{red} 0.02$^{~~}$} & {\color{red} 0.03$^{~~}$} & {\color{red} 0.02$^{~~}$}&&& {\color{red} 0.17$^{~~}$} & {\color{red} 0.14$^{~~}$} & {\color{red} 0.07$^{~~}$} & {\color{red} 0.09$^{~~}$} & {\color{red} 0.08$^{~~}$} & {\color{red} 0.02$^{~~}$} & {\color{red} 0.01$^{~~}$} & {\color{red} 0.02$^{~~}$} & {\color{red} 0.01$^{~~}$} \\
   &$G3$& {\color{red} 0.21$^{~~}$} & {\color{red} 0.11$^{~~}$} & {\color{red} 0.09$^{~~}$} & {\color{red} 0.11$^{~~}$} & {\color{red} 0.08$^{~~}$} & {\color{red} 0.02$^{~~}$} & {\color{red} 0.01$^{~~}$} & {\color{red} 0.01$^{~~}$} & {\color{red} 0.01$^{~~}$}&&& {\color{red} 0.16$^{~~}$} & {\color{red} 0.11$^{~~}$} & {\color{red} 0.09$^{~~}$} & {\color{red} 0.11$^{~~}$} & {\color{red} 0.09$^{~~}$} & {\color{red} 0.02$^{~~}$} & {\color{red} 0.01$^{~~}$} & {\color{red} 0.01$^{~~}$} & {\color{red} 0.01$^{~~}$} \\
   &$G4$& {\color{blue}-0.02$^{~~}$} & {\color{red} 0.01$^{~~}$} & {\color{blue}-0.02$^{~~}$} & {\color{red} 0.04$^{~~}$} & {\color{red} 0.04$^{~~}$} & {\color{red} 0.00$^{~~}$} & {\color{red} 0.00$^{~~}$} & {\color{red} 0.00$^{~~}$} & {\color{blue}-0.00$^{~~}$}&&& {\color{blue}-0.00$^{~~}$} & {\color{red} 0.01$^{~~}$} & {\color{blue}-0.01$^{~~}$} & {\color{red} 0.04$^{~~}$} & {\color{red} 0.03$^{~~}$} & {\color{blue}-0.00$^{~~}$} & {\color{blue}-0.00$^{~~}$} & {\color{blue}-0.00$^{~~}$} & {\color{blue}-0.00$^{~~}$} \\
   &$G5$& {\color{red} 0.11$^{~~}$} & {\color{red} 0.12$^{~~}$} & {\color{red} 0.07$^{~~}$} & {\color{red} 0.06$^{~~}$} & {\color{red} 0.04$^{~~}$} & {\color{red} 0.03$^{~~}$} & {\color{red} 0.03$^{~~}$} & {\color{red} 0.02$^{~~}$} & {\color{red} 0.01$^{~~}$}&&& {\color{red} 0.10$^{~~}$} & {\color{red} 0.12$^{~~}$} & {\color{red} 0.08$^{~~}$} & {\color{red} 0.06$^{~~}$} & {\color{red} 0.05$^{~~}$} & {\color{red} 0.03$^{~~}$} & {\color{red} 0.02$^{~~}$} & {\color{red} 0.02$^{~~}$} & {\color{red} 0.01$^{~~}$} \\
\hline
   $3$ &$G1$& {\color{red} \bf  0.26$^{**}$} & {\color{red} 0.13$^{~~}$} & {\color{red} 0.09$^{~~}$} & {\color{red} 0.13$^{~~}$} & {\color{red} 0.12$^{~~}$} & {\color{red} 0.08$^{~~}$} & {\color{red} 0.07$^{~~}$} & {\color{red} 0.06$^{~~}$} & {\color{red} 0.05$^{~~}$}&&& {\color{red} \bf  0.25$^{*~}$} & {\color{red} 0.16$^{~~}$} & {\color{red} 0.15$^{~~}$} & {\color{red} \bf  0.19$^{*~}$} & {\color{red} \bf  0.16$^{*~}$} & {\color{red} 0.12$^{~~}$} & {\color{red} 0.10$^{~~}$} & {\color{red} 0.10$^{~~}$} & {\color{red} 0.08$^{~~}$} \\
   &$G2$& {\color{red} \bf  0.26$^{**}$} & {\color{red} 0.14$^{~~}$} & {\color{red} 0.08$^{~~}$} & {\color{red} 0.10$^{~~}$} & {\color{red} 0.07$^{~~}$} & {\color{blue}-0.00$^{~~}$} & {\color{blue}-0.01$^{~~}$} & {\color{blue}-0.00$^{~~}$} & {\color{blue}-0.01$^{~~}$}&&& {\color{red} 0.17$^{~~}$} & {\color{red} 0.10$^{~~}$} & {\color{red} 0.09$^{~~}$} & {\color{red} 0.12$^{~~}$} & {\color{red} 0.09$^{~~}$} & {\color{red} 0.01$^{~~}$} & {\color{red} 0.00$^{~~}$} & {\color{red} 0.01$^{~~}$} & {\color{red} 0.00$^{~~}$} \\
   &$G3$& {\color{red} \bf  0.27$^{*~}$} & {\color{red} 0.17$^{~~}$} & {\color{red} 0.12$^{~~}$} & {\color{red} 0.14$^{~~}$} & {\color{red} 0.11$^{~~}$} & {\color{red} 0.02$^{~~}$} & {\color{red} 0.01$^{~~}$} & {\color{red} 0.01$^{~~}$} & {\color{red} 0.01$^{~~}$}&&& {\color{red} 0.15$^{~~}$} & {\color{red} 0.15$^{~~}$} & {\color{red} 0.14$^{~~}$} & {\color{red} 0.16$^{~~}$} & {\color{red} 0.12$^{~~}$} & {\color{red} 0.03$^{~~}$} & {\color{red} 0.02$^{~~}$} & {\color{red} 0.02$^{~~}$} & {\color{red} 0.02$^{~~}$} \\
   &$G4$& {\color{red} 0.03$^{~~}$} & {\color{red} 0.01$^{~~}$} & {\color{red} 0.01$^{~~}$} & {\color{red} 0.06$^{~~}$} & {\color{red} 0.07$^{~~}$} & {\color{red} 0.02$^{~~}$} & {\color{red} 0.01$^{~~}$} & {\color{red} 0.02$^{~~}$} & {\color{red} 0.01$^{~~}$}&&& {\color{red} 0.03$^{~~}$} & {\color{red} 0.04$^{~~}$} & {\color{red} 0.05$^{~~}$} & {\color{red} 0.08$^{~~}$} & {\color{red} 0.08$^{~~}$} & {\color{red} 0.02$^{~~}$} & {\color{red} 0.02$^{~~}$} & {\color{red} 0.02$^{~~}$} & {\color{red} 0.02$^{~~}$} \\
   &$G5$& {\color{red} 0.11$^{~~}$} & {\color{red} 0.09$^{~~}$} & {\color{red} 0.06$^{~~}$} & {\color{red} 0.05$^{~~}$} & {\color{red} 0.06$^{~~}$} & {\color{red} 0.03$^{~~}$} & {\color{red} 0.02$^{~~}$} & {\color{red} 0.01$^{~~}$} & {\color{red} 0.00$^{~~}$}&&& {\color{red} 0.18$^{~~}$} & {\color{red} 0.16$^{~~}$} & {\color{red} 0.09$^{~~}$} & {\color{red} 0.07$^{~~}$} & {\color{red} 0.07$^{~~}$} & {\color{red} 0.02$^{~~}$} & {\color{red} 0.01$^{~~}$} & {\color{red} 0.00$^{~~}$} & {\color{blue}-0.00$^{~~}$} \\
\hline
   $6$ &$G1$& {\color{red} \bf  0.32$^{**}$} & {\color{red} 0.17$^{~~}$} & {\color{red} 0.15$^{~~}$} & {\color{red} \bf  0.18$^{*~}$} & {\color{red} 0.14$^{~~}$} & {\color{red} 0.06$^{~~}$} & {\color{red} 0.06$^{~~}$} & {\color{red} 0.06$^{~~}$} & {\color{red} 0.04$^{~~}$}&&& {\color{red} \bf  0.23$^{*~}$} & {\color{red} 0.18$^{~~}$} & {\color{red} \bf  0.21$^{*~}$} & {\color{red} \bf  0.21$^{*~}$} & {\color{red} \bf  0.18$^{*~}$} & {\color{red} 0.12$^{~~}$} & {\color{red} 0.11$^{~~}$} & {\color{red} 0.11$^{~~}$} & {\color{red} 0.09$^{~~}$} \\
   &$G2$& {\color{red} 0.20$^{~~}$} & {\color{red} 0.14$^{~~}$} & {\color{red} 0.15$^{~~}$} & {\color{red} \bf  0.17$^{*~}$} & {\color{red} 0.14$^{~~}$} & {\color{red} 0.06$^{~~}$} & {\color{red} 0.05$^{~~}$} & {\color{red} 0.06$^{~~}$} & {\color{red} 0.05$^{~~}$}&&& {\color{red} 0.06$^{~~}$} & {\color{red} 0.10$^{~~}$} & {\color{red} 0.17$^{~~}$} & {\color{red} \bf  0.18$^{*~}$} & {\color{red} 0.14$^{~~}$} & {\color{red} 0.07$^{~~}$} & {\color{red} 0.06$^{~~}$} & {\color{red} 0.06$^{~~}$} & {\color{red} 0.05$^{~~}$} \\
   &$G3$& {\color{red} 0.22$^{~~}$} & {\color{red} 0.15$^{~~}$} & {\color{red} 0.16$^{~~}$} & {\color{red} 0.17$^{~~}$} & {\color{red} 0.12$^{~~}$} & {\color{red} 0.05$^{~~}$} & {\color{red} 0.03$^{~~}$} & {\color{red} 0.03$^{~~}$} & {\color{red} 0.02$^{~~}$}&&& {\color{red} 0.11$^{~~}$} & {\color{red} 0.10$^{~~}$} & {\color{red} 0.14$^{~~}$} & {\color{red} 0.11$^{~~}$} & {\color{red} 0.06$^{~~}$} & {\color{blue}-0.02$^{~~}$} & {\color{blue}-0.02$^{~~}$} & {\color{blue}-0.02$^{~~}$} & {\color{blue}-0.03$^{~~}$} \\
   &$G4$& {\color{red} 0.05$^{~~}$} & {\color{red} 0.01$^{~~}$} & {\color{red} 0.02$^{~~}$} & {\color{red} 0.05$^{~~}$} & {\color{red} 0.04$^{~~}$} & {\color{red} 0.02$^{~~}$} & {\color{blue}-0.00$^{~~}$} & {\color{blue}-0.00$^{~~}$} & {\color{blue}-0.01$^{~~}$}&&& {\color{blue}-0.06$^{~~}$} & {\color{blue}-0.01$^{~~}$} & {\color{red} 0.06$^{~~}$} & {\color{red} 0.06$^{~~}$} & {\color{red} 0.04$^{~~}$} & {\color{blue}-0.00$^{~~}$} & {\color{red} 0.00$^{~~}$} & {\color{red} 0.01$^{~~}$} & {\color{blue}-0.00$^{~~}$} \\
   &$G5$& {\color{red} 0.18$^{~~}$} & {\color{red} 0.14$^{~~}$} & {\color{red} 0.08$^{~~}$} & {\color{red} 0.06$^{~~}$} & {\color{red} 0.07$^{~~}$} & {\color{red} 0.04$^{~~}$} & {\color{red} 0.01$^{~~}$} & {\color{red} 0.01$^{~~}$} & {\color{blue}-0.00$^{~~}$}&&& {\color{red} 0.05$^{~~}$} & {\color{red} 0.07$^{~~}$} & {\color{red} 0.06$^{~~}$} & {\color{red} 0.05$^{~~}$} & {\color{red} 0.05$^{~~}$} & {\color{blue}-0.01$^{~~}$} & {\color{blue}-0.02$^{~~}$} & {\color{blue}-0.02$^{~~}$} & {\color{blue}-0.03$^{~~}$} \\
\hline
   $9$ &$G1$& {\color{red} 0.17$^{~~}$} & {\color{red} 0.14$^{~~}$} & {\color{red} 0.16$^{~~}$} & {\color{red} 0.16$^{~~}$} & {\color{red} 0.13$^{~~}$} & {\color{red} 0.05$^{~~}$} & {\color{red} 0.06$^{~~}$} & {\color{red} 0.06$^{~~}$} & {\color{red} 0.04$^{~~}$}&&& {\color{red} \bf  0.29$^{**}$} & {\color{red} \bf  0.23$^{*~}$} & {\color{red} \bf  0.22$^{*~}$} & {\color{red} \bf  0.19$^{*~}$} & {\color{red} \bf  0.18$^{*~}$} & {\color{red} 0.10$^{~~}$} & {\color{red} 0.11$^{~~}$} & {\color{red} 0.10$^{~~}$} & {\color{red} 0.08$^{~~}$} \\
   &$G2$& {\color{red} 0.06$^{~~}$} & {\color{red} 0.12$^{~~}$} & {\color{red} 0.13$^{~~}$} & {\color{red} 0.13$^{~~}$} & {\color{red} 0.10$^{~~}$} & {\color{red} 0.02$^{~~}$} & {\color{red} 0.02$^{~~}$} & {\color{red} 0.02$^{~~}$} & {\color{red} 0.01$^{~~}$}&&& {\color{red} 0.19$^{~~}$} & {\color{red} \bf  0.19$^{*~}$} & {\color{red} \bf  0.21$^{*~}$} & {\color{red} \bf  0.18$^{*~}$} & {\color{red} 0.15$^{~~}$} & {\color{red} 0.06$^{~~}$} & {\color{red} 0.05$^{~~}$} & {\color{red} 0.04$^{~~}$} & {\color{red} 0.03$^{~~}$} \\
   &$G3$& {\color{red} \bf  0.26$^{*~}$} & {\color{red} 0.21$^{~~}$} & {\color{red} 0.20$^{~~}$} & {\color{red} 0.17$^{~~}$} & {\color{red} 0.13$^{~~}$} & {\color{red} 0.06$^{~~}$} & {\color{red} 0.04$^{~~}$} & {\color{red} 0.04$^{~~}$} & {\color{red} 0.03$^{~~}$}&&& {\color{red} 0.19$^{~~}$} & {\color{red} 0.20$^{~~}$} & {\color{red} \bf  0.17$^{*~}$} & {\color{red} 0.10$^{~~}$} & {\color{red} 0.08$^{~~}$} & {\color{blue}-0.01$^{~~}$} & {\color{blue}-0.02$^{~~}$} & {\color{blue}-0.03$^{~~}$} & {\color{blue}-0.04$^{~~}$} \\
   &$G4$& {\color{blue}-0.01$^{~~}$} & {\color{blue}-0.01$^{~~}$} & {\color{blue}-0.00$^{~~}$} & {\color{blue}-0.01$^{~~}$} & {\color{blue}-0.02$^{~~}$} & {\color{blue}-0.04$^{~~}$} & {\color{blue}-0.05$^{~~}$} & {\color{blue}-0.05$^{~~}$} & {\color{blue}-0.06$^{~~}$}&&& {\color{red} 0.04$^{~~}$} & {\color{red} 0.13$^{~~}$} & {\color{red} 0.10$^{~~}$} & {\color{red} 0.07$^{~~}$} & {\color{red} 0.06$^{~~}$} & {\color{red} 0.01$^{~~}$} & {\color{blue}-0.00$^{~~}$} & {\color{blue}-0.00$^{~~}$} & {\color{blue}-0.02$^{~~}$} \\
   &$G5$& {\color{red} 0.08$^{~~}$} & {\color{red} 0.07$^{~~}$} & {\color{red} 0.04$^{~~}$} & {\color{red} 0.05$^{~~}$} & {\color{red} 0.05$^{~~}$} & {\color{blue}-0.00$^{~~}$} & {\color{blue}-0.03$^{~~}$} & {\color{blue}-0.04$^{~~}$} & {\color{blue}-0.05$^{~~}$}&&& {\color{red} 0.17$^{~~}$} & {\color{red} 0.12$^{~~}$} & {\color{red} 0.08$^{~~}$} & {\color{red} 0.07$^{~~}$} & {\color{red} 0.08$^{~~}$} & {\color{blue}-0.03$^{~~}$} & {\color{blue}-0.06$^{~~}$} & {\color{blue}-0.07$^{~~}$} & {\color{blue}-0.08$^{~~}$} \\
\hline
   $12$ &$G1$& {\color{red} 0.19$^{~~}$} & {\color{red} 0.14$^{~~}$} & {\color{red} 0.15$^{~~}$} & {\color{red} 0.13$^{~~}$} & {\color{red} 0.10$^{~~}$} & {\color{red} 0.04$^{~~}$} & {\color{red} 0.05$^{~~}$} & {\color{red} 0.05$^{~~}$} & {\color{red} 0.04$^{~~}$}&&& {\color{red} \bf  0.24$^{*~}$} & {\color{red} 0.18$^{~~}$} & {\color{red} 0.19$^{~~}$} & {\color{red} 0.17$^{~~}$} & {\color{red} 0.15$^{~~}$} & {\color{red} 0.08$^{~~}$} & {\color{red} 0.10$^{~~}$} & {\color{red} 0.10$^{~~}$} & {\color{red} 0.08$^{~~}$} \\
   &$G2$& {\color{red} 0.05$^{~~}$} & {\color{red} 0.11$^{~~}$} & {\color{red} 0.11$^{~~}$} & {\color{red} 0.10$^{~~}$} & {\color{red} 0.08$^{~~}$} & {\color{red} 0.01$^{~~}$} & {\color{red} 0.01$^{~~}$} & {\color{red} 0.00$^{~~}$} & {\color{blue}-0.01$^{~~}$}&&& {\color{red} \bf  0.20$^{*~}$} & {\color{red} \bf  0.17$^{*~}$} & {\color{red} 0.14$^{~~}$} & {\color{red} 0.11$^{~~}$} & {\color{red} 0.09$^{~~}$} & {\color{blue}-0.00$^{~~}$} & {\color{red} 0.01$^{~~}$} & {\color{red} 0.00$^{~~}$} & {\color{blue}-0.00$^{~~}$} \\
   &$G3$& {\color{red} 0.22$^{~~}$} & {\color{red} 0.19$^{~~}$} & {\color{red} 0.17$^{~~}$} & {\color{red} 0.13$^{~~}$} & {\color{red} 0.10$^{~~}$} & {\color{red} 0.03$^{~~}$} & {\color{red} 0.03$^{~~}$} & {\color{red} 0.02$^{~~}$} & {\color{red} 0.01$^{~~}$}&&& {\color{red} \bf  0.26$^{*~}$} & {\color{red} \bf  0.22$^{*~}$} & {\color{red} 0.18$^{~~}$} & {\color{red} 0.15$^{~~}$} & {\color{red} 0.12$^{~~}$} & {\color{red} 0.02$^{~~}$} & {\color{red} 0.03$^{~~}$} & {\color{red} 0.03$^{~~}$} & {\color{red} 0.02$^{~~}$} \\
   &$G4$& {\color{blue}-0.01$^{~~}$} & {\color{red} 0.03$^{~~}$} & {\color{blue}-0.00$^{~~}$} & {\color{blue}-0.02$^{~~}$} & {\color{blue}-0.03$^{~~}$} & {\color{blue}-0.05$^{~~}$} & {\color{blue}-0.06$^{~~}$} & {\color{blue}-0.06$^{~~}$} & {\color{blue}-0.07$^{~~}$}&&& {\color{red} 0.15$^{~~}$} & {\color{red} 0.14$^{~~}$} & {\color{red} 0.11$^{~~}$} & {\color{red} 0.09$^{~~}$} & {\color{red} 0.07$^{~~}$} & {\color{red} 0.03$^{~~}$} & {\color{red} 0.03$^{~~}$} & {\color{red} 0.03$^{~~}$} & {\color{red} 0.02$^{~~}$} \\
   &$G5$& {\color{red} 0.14$^{~~}$} & {\color{red} 0.12$^{~~}$} & {\color{red} 0.09$^{~~}$} & {\color{red} 0.09$^{~~}$} & {\color{red} 0.09$^{~~}$} & {\color{red} 0.02$^{~~}$} & {\color{blue}-0.01$^{~~}$} & {\color{blue}-0.02$^{~~}$} & {\color{blue}-0.03$^{~~}$}&&& {\color{red} 0.15$^{~~}$} & {\color{red} 0.12$^{~~}$} & {\color{red} 0.12$^{~~}$} & {\color{red} 0.12$^{~~}$} & {\color{red} 0.08$^{~~}$} & {\color{blue}-0.02$^{~~}$} & {\color{blue}-0.03$^{~~}$} & {\color{blue}-0.04$^{~~}$} & {\color{blue}-0.05$^{~~}$} \\
 \hline
   $24$ &$G1$& {\color{red} \bf  0.18$^{*~}$} & {\color{red} 0.15$^{~~}$} & {\color{red} 0.14$^{~~}$} & {\color{red} 0.13$^{~~}$} & {\color{red} 0.10$^{~~}$} & {\color{red} 0.05$^{~~}$} & {\color{red} 0.08$^{~~}$} & {\color{red} 0.08$^{~~}$} & {\color{red} 0.07$^{~~}$}&&& {\color{red} \bf  0.16$^{*~}$} & {\color{red} 0.12$^{~~}$} & {\color{red} 0.07$^{~~}$} & {\color{red} 0.06$^{~~}$} & {\color{red} 0.04$^{~~}$} & {\color{red} 0.10$^{~~}$} & {\color{red} \bf  0.12$^{*~}$} & {\color{red} 0.12$^{~~}$} & {\color{red} 0.11$^{~~}$} \\
   &$G2$& {\color{red} 0.02$^{~~}$} & {\color{red} 0.05$^{~~}$} & {\color{red} 0.03$^{~~}$} & {\color{red} 0.01$^{~~}$} & {\color{blue}-0.01$^{~~}$} & {\color{blue}-0.06$^{~~}$} & {\color{blue}-0.04$^{~~}$} & {\color{blue}-0.04$^{~~}$} & {\color{blue}-0.05$^{~~}$}&&& {\color{red} 0.05$^{~~}$} & {\color{red} 0.01$^{~~}$} & {\color{blue}-0.03$^{~~}$} & {\color{blue}-0.05$^{~~}$} & {\color{blue}-0.07$^{~~}$} & {\color{blue}-0.06$^{~~}$} & {\color{blue}-0.04$^{~~}$} & {\color{blue}-0.05$^{~~}$} & {\color{blue}-0.05$^{~~}$} \\
   &$G3$& {\color{red} 0.07$^{~~}$} & {\color{red} 0.07$^{~~}$} & {\color{red} 0.02$^{~~}$} & {\color{blue}-0.01$^{~~}$} & {\color{blue}-0.03$^{~~}$} & {\color{blue}-0.07$^{~~}$} & {\color{blue}-0.06$^{~~}$} & {\color{blue}-0.06$^{~~}$} & {\color{blue}-0.07$^{~~}$}&&& {\color{blue}-0.03$^{~~}$} & {\color{blue}-0.03$^{~~}$} & {\color{blue}-0.07$^{~~}$} & {\color{blue}-0.08$^{~~}$} & {\color{blue}-0.10$^{~~}$} & {\color{blue}-0.11$^{~~}$} & {\color{blue}-0.09$^{~~}$} & {\color{blue}-0.10$^{~~}$} & {\color{blue}-0.10$^{~~}$} \\
   &$G4$& {\color{red} 0.08$^{~~}$} & {\color{red} 0.10$^{~~}$} & {\color{red} 0.07$^{~~}$} & {\color{red} 0.05$^{~~}$} & {\color{red} 0.04$^{~~}$} & {\color{blue}-0.00$^{~~}$} & {\color{blue}-0.01$^{~~}$} & {\color{blue}-0.01$^{~~}$} & {\color{blue}-0.02$^{~~}$}&&& {\color{red} 0.02$^{~~}$} & {\color{blue}-0.01$^{~~}$} & {\color{blue}-0.01$^{~~}$} & {\color{blue}-0.02$^{~~}$} & {\color{blue}-0.04$^{~~}$} & {\color{blue}-0.04$^{~~}$} & {\color{blue}-0.02$^{~~}$} & {\color{blue}-0.03$^{~~}$} & {\color{blue}-0.04$^{~~}$} \\
   &$G5$& {\color{red} 0.14$^{~~}$} & {\color{red} 0.14$^{~~}$} & {\color{red} 0.13$^{~~}$} & {\color{red} 0.14$^{~~}$} & {\color{red} 0.12$^{~~}$} & {\color{red} 0.05$^{~~}$} & {\color{red} 0.03$^{~~}$} & {\color{red} 0.01$^{~~}$} & {\color{blue}-0.01$^{~~}$}&&& {\color{red} 0.13$^{~~}$} & {\color{red} 0.11$^{~~}$} & {\color{red} 0.10$^{~~}$} & {\color{red} 0.08$^{~~}$} & {\color{red} 0.04$^{~~}$} & {\color{blue}-0.01$^{~~}$} & {\color{blue}-0.02$^{~~}$} & {\color{blue}-0.04$^{~~}$} & {\color{blue}-0.04$^{~~}$} \\
\hline
   $36$ &$G1$& {\color{red} 0.12$^{~~}$} & {\color{red} 0.10$^{~~}$} & {\color{red} 0.09$^{~~}$} & {\color{red} 0.08$^{~~}$} & {\color{red} 0.07$^{~~}$} & {\color{red} 0.10$^{~~}$} & {\color{red} \bf  0.12$^{*~}$} & {\color{red} \bf  0.13$^{*~}$} & {\color{red} \bf  0.13$^{*~}$}&&& {\color{red} 0.10$^{~~}$} & {\color{red} 0.07$^{~~}$} & {\color{red} 0.08$^{~~}$} & {\color{red} 0.11$^{~~}$} & {\color{red} 0.13$^{~~}$} & {\color{red} \bf  0.19$^{*~}$} & {\color{red} \bf  0.20$^{**}$} & {\color{red} \bf  0.19$^{**}$} & {\color{red} \bf  0.19$^{*~}$} \\
   &$G2$& {\color{red} 0.01$^{~~}$} & {\color{blue}-0.02$^{~~}$} & {\color{blue}-0.03$^{~~}$} & {\color{blue}-0.01$^{~~}$} & {\color{blue}-0.03$^{~~}$} & {\color{blue}-0.07$^{~~}$} & {\color{blue}-0.05$^{~~}$} & {\color{blue}-0.05$^{~~}$} & {\color{blue}-0.05$^{~~}$}&&& {\color{blue}-0.11$^{~~}$} & {\color{blue}-0.10$^{~~}$} & {\color{blue}-0.09$^{~~}$} & {\color{blue}-0.09$^{~~}$} & {\color{blue}-0.09$^{~~}$} & {\color{blue}-0.05$^{~~}$} & {\color{blue}-0.04$^{~~}$} & {\color{blue}-0.05$^{~~}$} & {\color{blue}-0.05$^{~~}$} \\
   &$G3$& {\color{red} 0.01$^{~~}$} & {\color{blue}-0.02$^{~~}$} & {\color{blue}-0.03$^{~~}$} & {\color{blue}-0.01$^{~~}$} & {\color{blue}-0.02$^{~~}$} & {\color{blue}-0.06$^{~~}$} & {\color{blue}-0.05$^{~~}$} & {\color{blue}-0.05$^{~~}$} & {\color{blue}-0.06$^{~~}$}&&& {\color{blue}-0.07$^{~~}$} & {\color{blue}-0.07$^{~~}$} & {\color{blue}-0.07$^{~~}$} & {\color{blue}-0.08$^{~~}$} & {\color{blue}-0.08$^{~~}$} & {\color{blue}-0.06$^{~~}$} & {\color{blue}-0.07$^{~~}$} & {\color{blue}-0.07$^{~~}$} & {\color{blue}-0.06$^{~~}$} \\
   &$G4$& {\color{red} 0.06$^{~~}$} & {\color{red} 0.07$^{~~}$} & {\color{red} 0.06$^{~~}$} & {\color{red} 0.07$^{~~}$} & {\color{red} 0.05$^{~~}$} & {\color{red} 0.01$^{~~}$} & {\color{red} 0.01$^{~~}$} & {\color{blue}-0.00$^{~~}$} & {\color{blue}-0.01$^{~~}$}&&& {\color{blue}-0.01$^{~~}$} & {\color{blue}-0.02$^{~~}$} & {\color{blue}-0.03$^{~~}$} & {\color{blue}-0.05$^{~~}$} & {\color{blue}-0.04$^{~~}$} & {\color{blue}-0.02$^{~~}$} & {\color{blue}-0.03$^{~~}$} & {\color{blue}-0.04$^{~~}$} & {\color{blue}-0.04$^{~~}$} \\
   &$G5$& {\color{red} 0.10$^{~~}$} & {\color{red} 0.14$^{~~}$} & {\color{red} 0.14$^{~~}$} & {\color{red} 0.12$^{~~}$} & {\color{red} 0.10$^{~~}$} & {\color{red} 0.03$^{~~}$} & {\color{blue}-0.00$^{~~}$} & {\color{blue}-0.02$^{~~}$} & {\color{blue}-0.03$^{~~}$}&&& {\color{red} 0.07$^{~~}$} & {\color{red} 0.02$^{~~}$} & {\color{blue}-0.01$^{~~}$} & {\color{blue}-0.03$^{~~}$} & {\color{blue}-0.04$^{~~}$} & {\color{blue}-0.06$^{~~}$} & {\color{blue}-0.08$^{~~}$} & {\color{blue}-0.08$^{~~}$} & {\color{blue}-0.07$^{~~}$} \\
\hline
   $48$ &$G1$& {\color{red} 0.10$^{~~}$} & {\color{red} 0.05$^{~~}$} & {\color{red} 0.05$^{~~}$} & {\color{red} 0.06$^{~~}$} & {\color{red} 0.06$^{~~}$} & {\color{red} 0.10$^{~~}$} & {\color{red} \bf  0.12$^{*~}$} & {\color{red} \bf  0.12$^{*~}$} & {\color{red} \bf  0.12$^{*~}$}&&& {\color{red} 0.13$^{~~}$} & {\color{red} \bf  0.15$^{*~}$} & {\color{red} \bf  0.16$^{*~}$} & {\color{red} \bf  0.16$^{*~}$} & {\color{red} \bf  0.15$^{*~}$} & {\color{red} \bf  0.17$^{*~}$} & {\color{red} \bf  0.17$^{*~}$} & {\color{red} \bf  0.18$^{*~}$} & {\color{red} \bf  0.18$^{*~}$} \\
   &$G2$& {\color{red} 0.03$^{~~}$} & {\color{red} 0.02$^{~~}$} & {\color{blue}-0.01$^{~~}$} & {\color{blue}-0.02$^{~~}$} & {\color{blue}-0.03$^{~~}$} & {\color{blue}-0.04$^{~~}$} & {\color{blue}-0.03$^{~~}$} & {\color{blue}-0.03$^{~~}$} & {\color{blue}-0.02$^{~~}$}&&& {\color{blue}-0.03$^{~~}$} & {\color{blue}-0.03$^{~~}$} & {\color{blue}-0.04$^{~~}$} & {\color{blue}-0.04$^{~~}$} & {\color{blue}-0.04$^{~~}$} & {\color{blue}-0.01$^{~~}$} & {\color{blue}-0.00$^{~~}$} & {\color{red} 0.00$^{~~}$} & {\color{red} 0.01$^{~~}$} \\
   &$G3$& {\color{red} 0.02$^{~~}$} & {\color{red} 0.01$^{~~}$} & {\color{red} 0.01$^{~~}$} & {\color{red} 0.00$^{~~}$} & {\color{blue}-0.01$^{~~}$} & {\color{blue}-0.05$^{~~}$} & {\color{blue}-0.04$^{~~}$} & {\color{blue}-0.05$^{~~}$} & {\color{blue}-0.04$^{~~}$}&&& {\color{blue}-0.04$^{~~}$} & {\color{blue}-0.07$^{~~}$} & {\color{blue}-0.08$^{~~}$} & {\color{blue}-0.07$^{~~}$} & {\color{blue}-0.08$^{~~}$} & {\color{blue}-0.05$^{~~}$} & {\color{blue}-0.05$^{~~}$} & {\color{blue}-0.05$^{~~}$} & {\color{blue}-0.03$^{~~}$} \\
   &$G4$& {\color{red} 0.01$^{~~}$} & {\color{red} 0.03$^{~~}$} & {\color{red} 0.04$^{~~}$} & {\color{red} 0.04$^{~~}$} & {\color{red} 0.02$^{~~}$} & {\color{blue}-0.01$^{~~}$} & {\color{blue}-0.01$^{~~}$} & {\color{blue}-0.03$^{~~}$} & {\color{blue}-0.03$^{~~}$}&&& {\color{blue}-0.02$^{~~}$} & {\color{blue}-0.04$^{~~}$} & {\color{blue}-0.06$^{~~}$} & {\color{blue}-0.07$^{~~}$} & {\color{blue}-0.08$^{~~}$} & {\color{blue}-0.07$^{~~}$} & {\color{blue}-0.08$^{~~}$} & {\color{blue}-0.08$^{~~}$} & {\color{blue}-0.06$^{~~}$} \\
   &$G5$& {\color{red} 0.09$^{~~}$} & {\color{red} 0.11$^{~~}$} & {\color{red} 0.11$^{~~}$} & {\color{red} 0.09$^{~~}$} & {\color{red} 0.07$^{~~}$} & {\color{red} 0.01$^{~~}$} & {\color{red} 0.00$^{~~}$} & {\color{blue}-0.02$^{~~}$} & {\color{blue}-0.02$^{~~}$}&&& {\color{red} 0.01$^{~~}$} & {\color{blue}-0.00$^{~~}$} & {\color{blue}-0.02$^{~~}$} & {\color{blue}-0.05$^{~~}$} & {\color{blue}-0.05$^{~~}$} & {\color{blue}-0.06$^{~~}$} & {\color{blue}-0.08$^{~~}$} & {\color{blue}-0.08$^{~~}$} & {\color{blue}-0.06$^{~~}$} \\
\hline
   $60$ &$G1$& {\color{red} 0.09$^{~~}$} & {\color{red} 0.05$^{~~}$} & {\color{red} 0.05$^{~~}$} & {\color{red} 0.04$^{~~}$} & {\color{red} 0.04$^{~~}$} & {\color{red} 0.09$^{~~}$} & {\color{red} \bf  0.11$^{*~}$} & {\color{red} \bf  0.12$^{*~}$} & {\color{red} \bf  0.12$^{*~}$}&&& {\color{red} \bf  0.16$^{*~}$} & {\color{red} \bf  0.16$^{*~}$} & {\color{red} \bf  0.16$^{*~}$} & {\color{red} \bf  0.16$^{*~}$} & {\color{red} \bf  0.16$^{*~}$} & {\color{red} \bf  0.16$^{*~}$} & {\color{red} \bf  0.16$^{*~}$} & {\color{red} \bf  0.17$^{*~}$} & {\color{red} \bf  0.16$^{*~}$} \\
   &$G2$& {\color{red} 0.01$^{~~}$} & {\color{blue}-0.02$^{~~}$} & {\color{blue}-0.05$^{~~}$} & {\color{blue}-0.06$^{~~}$} & {\color{blue}-0.07$^{~~}$} & {\color{blue}-0.06$^{~~}$} & {\color{blue}-0.04$^{~~}$} & {\color{blue}-0.04$^{~~}$} & {\color{blue}-0.03$^{~~}$}&&& {\color{blue}-0.05$^{~~}$} & {\color{blue}-0.05$^{~~}$} & {\color{blue}-0.04$^{~~}$} & {\color{blue}-0.04$^{~~}$} & {\color{blue}-0.04$^{~~}$} & {\color{blue}-0.02$^{~~}$} & {\color{blue}-0.00$^{~~}$} & {\color{red} 0.01$^{~~}$} & {\color{red} 0.01$^{~~}$} \\
   &$G3$& {\color{blue}-0.01$^{~~}$} & {\color{blue}-0.01$^{~~}$} & {\color{blue}-0.03$^{~~}$} & {\color{blue}-0.03$^{~~}$} & {\color{blue}-0.05$^{~~}$} & {\color{blue}-0.06$^{~~}$} & {\color{blue}-0.06$^{~~}$} & {\color{blue}-0.06$^{~~}$} & {\color{blue}-0.05$^{~~}$}&&& {\color{blue}-0.10$^{~~}$} & {\color{blue}-0.10$^{~~}$} & {\color{blue}-0.10$^{~~}$} & {\color{blue} \bf -0.11$^{*~}$} & {\color{blue} \bf -0.10$^{*~}$} & {\color{blue}-0.06$^{~~}$} & {\color{blue}-0.05$^{~~}$} & {\color{blue}-0.04$^{~~}$} & {\color{blue}-0.03$^{~~}$} \\
   &$G4$& {\color{red} 0.02$^{~~}$} & {\color{red} 0.00$^{~~}$} & {\color{red} 0.00$^{~~}$} & {\color{red} 0.00$^{~~}$} & {\color{blue}-0.00$^{~~}$} & {\color{blue}-0.01$^{~~}$} & {\color{blue}-0.03$^{~~}$} & {\color{blue}-0.05$^{~~}$} & {\color{blue}-0.04$^{~~}$}&&& {\color{blue}-0.07$^{~~}$} & {\color{blue}-0.07$^{~~}$} & {\color{blue}-0.08$^{~~}$} & {\color{blue}-0.09$^{~~}$} & {\color{blue}-0.10$^{~~}$} & {\color{blue}-0.08$^{~~}$} & {\color{blue}-0.08$^{~~}$} & {\color{blue}-0.08$^{~~}$} & {\color{blue}-0.06$^{~~}$} \\
   &$G5$& {\color{red} 0.10$^{~~}$} & {\color{red} 0.10$^{~~}$} & {\color{red} 0.09$^{~~}$} & {\color{red} 0.07$^{~~}$} & {\color{red} 0.06$^{~~}$} & {\color{red} 0.01$^{~~}$} & {\color{red} 0.00$^{~~}$} & {\color{blue}-0.01$^{~~}$} & {\color{blue}-0.00$^{~~}$}&&& {\color{red} 0.02$^{~~}$} & {\color{blue}-0.00$^{~~}$} & {\color{blue}-0.03$^{~~}$} & {\color{blue}-0.04$^{~~}$} & {\color{blue}-0.05$^{~~}$} & {\color{blue}-0.07$^{~~}$} & {\color{blue}-0.09$^{~~}$} & {\color{blue}-0.08$^{~~}$} & {\color{blue}-0.06$^{~~}$} \\
   \hline
   \end{tabular}
   \label{TB:ChinaGroupIndex:Closing Price}
\end{table}

It is apparent that look-back ($J$) and holding horizons ($K$), are somewhat negatively correlated with the performance of the TSMOM strategies. We observe that prolonging look-back and holding horizons corresponds to the increasing number of statistically significant TSCON strategies, especially when $J$, $K$ $\geq 24$, which is consistent with the previous findings. Concerning the closing price factor in Table \ref{TB:ChinaGroupIndex:Closing Price}, all the TSMOM strategies for group G1 have positive returns and the majority of the TSMOM strategies with statistical significance are concentrated in group G1, corresponding to the highest value of closing price, meanwhile the TSMOM  strategies for group G1 generally outperform those for group G5.
Concerning the adjusted price factor, we observe that most of the significant strategies are concentrated in group G1 and all returns are positive. Furthermore, the strategies for group G1 outperform those for group G5. Based on these findings, we argue that higher closing price and adjusted price could lead to a better performance of the TSMOM strategies.
For the market value factor (based on both outstanding shares and all shares), there exists a weak TSMOM effect, and the strategies with statistical significance are dispersed rather than concentrated in certain groups. The results about the TSMOM effect are mixed. We further find that the horizon of $J$ has a more evident impact on the relative performance of the TSMOM strategies based on different groups. Specifically, the groups with higher market value have better performance than those with lower market value when $J < 6$, while it turns out to be the opposite with $J$ keeping prolonging. The TSCON effect prevails over the TSMOM effect, especially under the signaling method of MOP. %
As for the turnover rate factor, although there are a few TSMOM strategies indicating the weak TSMOM effect, the performance of the TSMOM strategies is seemingly not sensitive to the turnover rate. On the other hand, we observe that the TSMOM strategies for group G5 have the better performance when look-back period $J$ is short-term. And also, it is stronger TSCON effect under the signaling method of MOP.
With respect to the trading volume factor, there are some significant TSMOM strategies, the majority of which are concentrated in group G5, corresponding to the lowest value of trading volume. Apparently, most of the TSMOM strategies outperform those for group G1. Also, it is a robust finding that there is a pronounced long-term TSCON effect, especially with the signaling method of MOP.
In addition, it is also noticeable that the TSMOM strategies are more profitable with short-term holding period $K$, especially when $K=1$, independent on whether the trading signaling method is HL or MOP.

For the clarity of observations, we additionally construct the following linear regression with inclusion of dummy variable for stock group to capture the relation of the average excess return to the TSMOM strategy and aforementioned firm-specific factors:
\begin{equation}
  ER({\mathrm{TSMOM}})=\alpha+\beta_{J}J+\beta_{K}K+g_{2}D_2+g_{3}D_3+g_{4}D_4+g_{5}D_5+\varepsilon,
  ~~~\varepsilon \sim N(0,\sigma^{2}_{\varepsilon}),
\label{Eq:GroupLR}
\end{equation}
where $ER$ denotes the annualized average monthly excess return to the TSMOM strategy, and $D_2$ to $D_5$ are the dummy variables with the value being unity if the TSMOM strategy is for its specific stock group and zero otherwise. The results are presented in Table \ref{TB:ChinaGroupIndexRegression} for different firm-specific characteristics mentioned above.

\begin{table}[!th]
  \caption{This table reports the results based on the linear regression model, $ER({\mathrm{TSMOM}})=\alpha+\beta_{J}J+\beta_{K}K+g_{2}D_2+g_{3}D_3+g_{4}D_4+g_{5}D_5$. $ER$ is the annualized monthly excess return to TSMOM strategy with various look-back and holding periods $J$ and $K$. The dummy variable $D$ takes the value of unity (zero) when the TSMOM strategy is applied to the corresponding stock group. The numbers in parentheses are the $t$-statistics. The superscripts * and ** denote the significance at 5\% and 1\% levels, respectively.}
\medskip
\centering
\small
   \begin{tabular}{ccccccccccccccc}
   \hline
     && $\alpha$ && $\beta_{J}$ &&  $\beta_{K}$ && $g_{2}$ && $g_{3}$ && $g_{4}$ && $g_{5}$ \\
   \hline
   \multicolumn{15}{l}{\textit{Panel A1: Closing price}} \\
   HL && 0.1643 && -0.0011 && -0.0017 && -0.0717 && -0.0581 && -0.0961 && -0.0485  \\
    && {\bf (25.12)$^{**}$} && {\bf (-9.27)$^{**}$} && {\bf (-14.23)$^{**}$} &&  {\bf (-9.43)$^{**}$} && {\bf (-7.64)$^{**}$} && {\bf (-12.65)$^{**}$} && {\bf (-6.39)$^{**}$} \\
   MOP && 0.2137 && -0.0019 && -0.0014 && -0.1106 && -0.1306 && -0.1444 && -0.1263  \\
    && {\bf (27.83)$^{**}$} && {\bf (-13.30)$^{**}$} && {\bf (-9.72)$^{**}$} &&  {\bf (-12.40)$^{**}$} && {\bf (-14.65)$^{**}$} && {\bf (-16.20)$^{**}$} && {\bf (-14.16)$^{**}$} \\
\hline
   \multicolumn{15}{l}{\textit{Panel A2: Adjusted price}} \\
   HL && 0.2460 && -0.0008 && -0.0020 && -0.1730 && -0.1853 && -0.1452 && -0.1194  \\
    && {\bf (37.76)$^{**}$} && {\bf (-6.32)$^{**}$} && {\bf (-16.87)$^{**}$} &&  {\bf (-22.87)$^{**}$} && {\bf (-24.49)$^{**}$} && {\bf (-19.19)$^{**}$} && {\bf (-15.79)$^{**}$} \\
   MOP && 0.2932 && -0.0016 && -0.0016 && -0.2078 && -0.2425 && -0.2344 && -0.2085  \\
    && {\bf (41.03)$^{**}$} && {\bf (-12.31)$^{**}$} && {\bf (-12.50)$^{**}$} &&  {\bf (-25.04)$^{**}$} && {\bf (-29.22)$^{**}$} && {\bf (-28.25)$^{**}$} && {\bf (-25.13)$^{**}$} \\
\hline
   \multicolumn{15}{l}{\textit{Panel A3: Market value (based on outstanding shares)}} \\
   HL && 0.0667 && -0.0012 && -0.0017 && -0.0097 && -0.0075 && 0.0726 && 0.0393  \\
    && {\bf (11.85)$^{**}$} && {\bf (-11.66)$^{**}$} && {\bf (-16.68)$^{**}$} &&  {(-1.49)$^{~~}$} && {(-1.15)$^{~~}$} && {\bf (11.11)$^{**}$} && {\bf ( 6.02)$^{**}$} \\
   MOP && 0.0806 && -0.0022 && -0.0014 && -0.0273 && -0.0397 && 0.0343 && 0.0238  \\
    && {\bf (11.74)$^{**}$} && {\bf (-17.68)$^{**}$} && {\bf (-10.97)$^{**}$} &&  {\bf (-3.43)$^{**}$} && {\bf (-4.98)$^{**}$} && {\bf ( 4.30)$^{**}$} && {\bf ( 2.98)$^{**}$} \\
\hline
   \multicolumn{15}{l}{\textit{Panel A4: Market value (based on all shares)}} \\
   HL && 0.0475 && -0.0010 && -0.0016 && 0.0012 && 0.0128 && 0.0352 && 0.0484  \\
    && {\bf (11.42)$^{**}$} && {\bf (-13.59)$^{**}$} && {\bf (-20.85)$^{**}$} &&  {( 0.24)$^{~~}$} && {\bf ( 2.64)$^{**}$} && {\bf ( 7.29)$^{**}$} && {\bf (10.03)$^{**}$} \\
   MOP && 0.0721 && -0.0020 && -0.0013 && -0.0280 && -0.0269 && -0.0173 && 0.0202  \\
    && {\bf (14.27)$^{**}$} && {\bf (-21.38)$^{**}$} && {\bf (-13.79)$^{**}$} &&  {\bf (-4.77)$^{**}$} && {\bf (-4.58)$^{**}$} && {\bf (-2.95)$^{**}$} && {\bf ( 3.44)$^{**}$} \\
\hline
   \multicolumn{15}{l}{\textit{Panel A5: Turnover rate}} \\
   HL && 0.0598 && -0.0011 && -0.0016 && 0.0200 && 0.0433 && 0.0404 && 0.0511  \\
    && {\bf (11.07)$^{**}$} && {\bf (-10.87)$^{**}$} && {\bf (-16.53)$^{**}$} &&  {\bf ( 3.19)$^{**}$} && {\bf ( 6.89)$^{**}$} && {\bf ( 6.43)$^{**}$} && {\bf ( 8.14)$^{**}$} \\
   MOP && 0.1038 && -0.0020 && -0.0014 && -0.0469 && -0.0380 && -0.0345 && -0.0130  \\
    && {\bf (14.86)$^{**}$} && {\bf (-16.04)$^{**}$} && {\bf (-11.02)$^{**}$} &&  {\bf (-5.78)$^{**}$} && {\bf (-4.69)$^{**}$} && {\bf (-4.26)$^{**}$} && {(-1.60)$^{~~}$} \\
\hline
   \multicolumn{15}{l}{\textit{Panel A6: Trading volume}} \\
   HL && 0.0554 && -0.0014 && -0.0017 && 0.0103 && 0.0393 && 0.0514 && 0.1312  \\
    && {\bf (10.51)$^{**}$} && {\bf (-14.17)$^{**}$} && {\bf (-17.54)$^{**}$} &&  {( 1.68)$^{~~}$} && {\bf ( 6.42)$^{**}$} && {\bf ( 8.40)$^{**}$} && {\bf (21.45)$^{**}$} \\
   MOP && 0.0715 && -0.0025 && -0.0013 && -0.0217 && -0.0171 && 0.0001 && 0.0907  \\
    && {\bf (10.75)$^{**}$} && {\bf (-20.75)$^{**}$} && {\bf (-10.39)$^{**}$} &&  {\bf (-2.81)$^{**}$} && {\bf (-2.21)$^{*~}$} && {( 0.01)$^{~~}$} && {\bf (11.75)$^{**}$} \\
   \hline
   \end{tabular}
   \label{TB:ChinaGroupIndexRegression}
\end{table}

It is obvious and statistically significant that the look-back and holding horizons are negatively correlated with average excess return to the TSMOM strategy, irrespective of the firm-specific characteristics, which is consistent with the findings above. This also implies that the TSMOM effect would be more significant with short-term look-back and holding periods. In addition, the impact of $J$ and $K$ on the TSMOM profitability is symmetrical and dependent on different trading signaling methods. As for the results of closing price and adjusted price, $K$ has a stronger impact on the profitability of the TSMOM strategies than $J$ does. Nevertheless, as for the results of market value, turnover rate, and trading volume, the findings based on the HL method are consistent that $K$ has a more powerful influence than $J$, while it turns out to be the opposite under the MOP framework. In sum, short-term (long-term) look-back and holding periods give rise to better performance of the TSMOM strategies and therefore intensifies the TSMOM effect or weakens the TSCON effect.

We then pay more attention to the relation of the ranking groups of each characteristics and the profitability of the TSMOM strategies. As for the closing price factor in Panel A1 of Table \ref{TB:ChinaGroupIndexRegression}, the coefficients of dummy variables for various groups are significantly negative, indicating that the average monthly excess return of strategies for G1 is higher than the rest. Moreover, the profitability presents a downward trend from G1 to G4. Despite of a slightly increase from G4 to G5, the average excess return of strategies for G5 is still significantly lower than that for G1.
Similarly, in Panel A2 of Table \ref{TB:ChinaGroupIndexRegression}, in regard to the adjusted price factor, the coefficients of dummy variables for different groups are also significantly negative. Therefore, the average monthly excess returns to the strategies for G1 group are higher than the rest. We also observe that the profitability has a decline tendency from G1 to G3. Although it turns out to increase from G3 to G5, the average excess returns of strategies for G5 are still significantly lower than those for G1. Roughly, higher closing and adjusted prices may result in higher (lower) excess return to the TSMOM (TSCON) strategies. That is, the TSMOM (TSCON) effect would be intensified (weakened).
Our finding is in accordance with the results on the cross-sectional momentum effect in the Chinese market \cite{Wang-Wang-Xu-Duan-2006-cnRFEI}.

Regarding the results of market value in Panel A4 of Table \ref{TB:ChinaGroupIndexRegression}, it is obvious that the respective performance of strategies for G4 and G5 are significantly better than those for G1, and also better than the cases of G2 and G3. As to the results based on the outstanding shares, there is the downward trend of the profitability from G1 to G3, whereas the average excess returns turn out to increase from G3 to G4, and a slightly decrease happens from G4 to G5. In a word, the strategies for G4 and G5 have better performance than the other three groups. This finding is also robust to different trading signaling methods (HL and MOP). In contrast, the results in Panel A3 of Table \ref{TB:ChinaGroupIndexRegression} based on all shares largely depend on the trading signaling methods. Specifically, under HL, the average excess return of the TSMOM strategies monotonically increases from G1 to G5, while under MOP, after a decrease trend from G1 to G2, there is also a monotonically increase trend from G2 to G5. In brief, a lower market value, based on both of outstanding and all shares, leads to more profitable TSMOM strategies.

As show in Panel A5 of Table \ref{TB:ChinaGroupIndexRegression}, the results of turnover rate are mixed. Under HL, there is a seemingly  monotonically increasing trend of profitability from G1 to G5, while under MOP, the average excess returns firstly decrease from G1 to G2 whereas increase monotonically from G2 to G5. As a result, we observe that the strategies for G5 has the best average performance under HL, while those for G1 has the best performance under MOP. As for the results of the trading volume factor in Panel A6 of Table \ref{TB:ChinaGroupIndexRegression}, our findings are similar to those of market value based on all shares. Specifically, under HL, the average excess return of the TSMOM strategies monotonically increase from G1 to G5, while under MOP, after a decrease trend from G1 to G2, there is also a monotonically increase trend from G2 to G5. Finally, the strategies for G5 has the best average performance, implying that lower trading volumes contribute to improving the profitability of the TSMOM strategies.

We can now draw the following conclusions. Firstly, in the short term, the TSMOM strategies in terms of the higher price (including the closing and adjusted price), higher market value and lower trading volume are more profitable and more likely to be statistical significant. In addition, the TSCON effect still prevails over the TSMOM effect, especially when the look-back period and the holding horizon are longer. It is also should be noted that our findings are consistent with those associated with cross-sectional momentum effect in the Chinese stock market. In comparison, the cross-sectional momentum portfolios based on higher price, higher market value and lower turnover rate are more profitable \cite{Wang-Wang-Xu-Duan-2006-cnRFEI}.

\subsection{The performance of stock groups based on industrial sectors}

A strand of literature on the cross-sectional momentum effect also finds that the presence of the industry momentum effect is ubiquitous in many markets. The momentum return is largely explained by the industry momentum in the US market \cite{Moskowitz-Grinblatt-1999-JF}. Accordingly, we also intend to evaluate the performance of the TSMOM strategies based on industrial sectors. Likewise, we divide the individual stocks into decile groups according to industrial sectors.
The $J$-$K$ TSMOM strategy is applied to the time series of equally-weighted average monthly return of each stock group. This method is to buy or sell all stocks from the same industrial sector according to the trading signals. In this manner, the stock groups can be regarded as various industry indices. An alternative method is to apply the $J$-$K$ TSMOM strategy to the individual stocks from certain decile group. Thus, the individual stocks from a same industrial sector are entered into the long or short positions according to their respective trading signal and we take an average of the average monthly excess returns of individual stocks in a certain industrial sector to conduct the adjusted $t$ test \cite{Newey-West-1987-Em}.

We follow the grouping rules for industrial sectors by the China Securities Index Co. Ltd. and the individual stocks are grouped into ten categories of industrial sectors, including Energy, Materials, Industry, Discretionary Consumption, Essential Consumption, Medical and Health, Finance and Estate, Technology, Telecommunications, and Public Utilities.
Table \ref{TB:ChinaIndustrySector:AlterConsume} presents the results for the sector of Discretionary Consumption as an example.

\setlength\tabcolsep{0.7pt}
\begin{table}[!ht]
  \caption{(Color online) The annualized monthly excess returns to TSMOM strategy applied to stock group of \textit{Discretionary Consumption}. Panel A is for the implementation of TSMOM to stock group ``index' based on industrial sectors and Panel B is for individual stocks within each stock group. The look-back and holding periods are $J$ and $K$. The superscripts * and ** denote the significance at 5\% and 1\% levels, respectively.}
\medskip
\centering
\footnotesize
   \begin{tabular}{lccccccccccccccccccccc}
   \hline
   && \multicolumn{9}{c}{HL}  &&& \multicolumn{9}{c}{MOP} \\
   \cline{3-11} \cline{14-22}
    $J$ && $K=1$ &3 & 6 & 9 & 12 & 24 & 36 & 48 & 60 &&& 1 & 3 & 6 & 9 & 12 & 24 & 36 & 48 & 60\\
   \hline
   \multicolumn{22}{l}{\textit{Panel A: Industrial sector index}} \\
   $1$ && {\color{red} 0.17$^{~~}$} & {\color{red} \bf  0.16$^{*~}$} & {\color{red} 0.13$^{~~}$} & {\color{red} 0.11$^{~~}$} & {\color{red} 0.09$^{~~}$} & {\color{red} 0.02$^{~~}$} & {\color{blue}-0.00$^{~~}$} & {\color{red} 0.01$^{~~}$} & {\color{red} 0.00$^{~~}$}&&& {\color{red} 0.19$^{~~}$} & {\color{red} \bf  0.17$^{*~}$} & {\color{red} 0.14$^{~~}$} & {\color{red} \bf  0.13$^{*~}$} & {\color{red} 0.10$^{~~}$} & {\color{red} 0.03$^{~~}$} & {\color{red} 0.00$^{~~}$} & {\color{red} 0.01$^{~~}$} & {\color{red} 0.01$^{~~}$} \\
   $3$ && {\color{red} \bf  0.23$^{*~}$} & {\color{red} 0.13$^{~~}$} & {\color{red} 0.10$^{~~}$} & {\color{red} 0.12$^{~~}$} & {\color{red} 0.11$^{~~}$} & {\color{red} 0.02$^{~~}$} & {\color{blue}-0.00$^{~~}$} & {\color{red} 0.00$^{~~}$} & {\color{blue}-0.00$^{~~}$}&&& {\color{red} 0.19$^{~~}$} & {\color{red} 0.13$^{~~}$} & {\color{red} 0.11$^{~~}$} & {\color{red} 0.13$^{~~}$} & {\color{red} 0.10$^{~~}$} & {\color{red} 0.01$^{~~}$} & {\color{blue}-0.01$^{~~}$} & {\color{red} 0.00$^{~~}$} & {\color{blue}-0.01$^{~~}$} \\
   $6$ && {\color{red} \bf  0.23$^{*~}$} & {\color{red} 0.12$^{~~}$} & {\color{red} 0.14$^{~~}$} & {\color{red} 0.15$^{~~}$} & {\color{red} 0.12$^{~~}$} & {\color{red} 0.02$^{~~}$} & {\color{blue}-0.00$^{~~}$} & {\color{red} 0.00$^{~~}$} & {\color{blue}-0.00$^{~~}$}&&& {\color{red} 0.15$^{~~}$} & {\color{red} 0.09$^{~~}$} & {\color{red} 0.14$^{~~}$} & {\color{red} 0.13$^{~~}$} & {\color{red} 0.09$^{~~}$} & {\color{blue}-0.00$^{~~}$} & {\color{blue}-0.01$^{~~}$} & {\color{blue}-0.01$^{~~}$} & {\color{blue}-0.02$^{~~}$} \\
   $9$ && {\color{red} 0.15$^{~~}$} & {\color{red} 0.13$^{~~}$} & {\color{red} 0.13$^{~~}$} & {\color{red} 0.12$^{~~}$} & {\color{red} 0.10$^{~~}$} & {\color{red} 0.00$^{~~}$} & {\color{blue}-0.01$^{~~}$} & {\color{blue}-0.01$^{~~}$} & {\color{blue}-0.02$^{~~}$}&&& {\color{red} \bf  0.25$^{*~}$} & {\color{red} 0.18$^{~~}$} & {\color{red} \bf  0.19$^{*~}$} & {\color{red} 0.15$^{~~}$} & {\color{red} 0.13$^{~~}$} & {\color{blue}-0.00$^{~~}$} & {\color{blue}-0.01$^{~~}$} & {\color{blue}-0.01$^{~~}$} & {\color{blue}-0.02$^{~~}$} \\
   $12$ && {\color{red} 0.14$^{~~}$} & {\color{red} 0.13$^{~~}$} & {\color{red} 0.12$^{~~}$} & {\color{red} 0.09$^{~~}$} & {\color{red} 0.07$^{~~}$} & {\color{blue}-0.02$^{~~}$} & {\color{blue}-0.01$^{~~}$} & {\color{blue}-0.01$^{~~}$} & {\color{blue}-0.03$^{~~}$}&&& {\color{red} 0.19$^{~~}$} & {\color{red} 0.15$^{~~}$} & {\color{red} 0.13$^{~~}$} & {\color{red} 0.11$^{~~}$} & {\color{red} 0.09$^{~~}$} & {\color{blue}-0.03$^{~~}$} & {\color{blue}-0.02$^{~~}$} & {\color{blue}-0.02$^{~~}$} & {\color{blue}-0.03$^{~~}$} \\
   $24$ && {\color{red} 0.14$^{~~}$} & {\color{red} 0.12$^{~~}$} & {\color{red} 0.08$^{~~}$} & {\color{red} 0.04$^{~~}$} & {\color{red} 0.01$^{~~}$} & {\color{blue}-0.04$^{~~}$} & {\color{blue}-0.03$^{~~}$} & {\color{blue}-0.02$^{~~}$} & {\color{blue}-0.03$^{~~}$}&&& {\color{red} 0.05$^{~~}$} & {\color{red} 0.02$^{~~}$} & {\color{blue}-0.02$^{~~}$} & {\color{blue}-0.05$^{~~}$} & {\color{blue}-0.08$^{~~}$} & {\color{blue}-0.08$^{~~}$} & {\color{blue}-0.06$^{~~}$} & {\color{blue}-0.06$^{~~}$} & {\color{blue}-0.06$^{~~}$} \\
   $36$ && {\color{red} 0.07$^{~~}$} & {\color{red} 0.03$^{~~}$} & {\color{red} 0.04$^{~~}$} & {\color{red} 0.04$^{~~}$} & {\color{red} 0.03$^{~~}$} & {\color{blue}-0.01$^{~~}$} & {\color{red} 0.01$^{~~}$} & {\color{red} 0.01$^{~~}$} & {\color{red} 0.00$^{~~}$}&&& {\color{blue}-0.00$^{~~}$} & {\color{blue}-0.01$^{~~}$} & {\color{blue}-0.02$^{~~}$} & {\color{blue}-0.03$^{~~}$} & {\color{blue}-0.03$^{~~}$} & {\color{red} 0.01$^{~~}$} & {\color{red} 0.00$^{~~}$} & {\color{blue}-0.01$^{~~}$} & {\color{red} 0.01$^{~~}$} \\
   $48$ && {\color{red} 0.05$^{~~}$} & {\color{red} 0.01$^{~~}$} & {\color{red} 0.01$^{~~}$} & {\color{red} 0.00$^{~~}$} & {\color{blue}-0.02$^{~~}$} & {\color{blue}-0.04$^{~~}$} & {\color{blue}-0.03$^{~~}$} & {\color{blue}-0.04$^{~~}$} & {\color{blue}-0.03$^{~~}$}&&& {\color{blue}-0.06$^{~~}$} & {\color{blue}-0.07$^{~~}$} & {\color{blue}-0.07$^{~~}$} & {\color{blue}-0.06$^{~~}$} & {\color{blue}-0.06$^{~~}$} & {\color{blue}-0.04$^{~~}$} & {\color{blue}-0.03$^{~~}$} & {\color{blue}-0.03$^{~~}$} & {\color{blue}-0.02$^{~~}$} \\
   $60$ && {\color{red} 0.00$^{~~}$} & {\color{blue}-0.03$^{~~}$} & {\color{blue}-0.04$^{~~}$} & {\color{blue}-0.03$^{~~}$} & {\color{blue}-0.04$^{~~}$} & {\color{blue}-0.04$^{~~}$} & {\color{blue}-0.04$^{~~}$} & {\color{blue}-0.05$^{~~}$} & {\color{blue}-0.04$^{~~}$}&&& {\color{blue}-0.09$^{~~}$} & {\color{blue}-0.08$^{~~}$} & {\color{blue}-0.08$^{~~}$} & {\color{blue}-0.08$^{~~}$} & {\color{blue}-0.08$^{~~}$} & {\color{blue}-0.05$^{~~}$} & {\color{blue}-0.03$^{~~}$} & {\color{blue}-0.02$^{~~}$} & {\color{blue}-0.01$^{~~}$} \\
\hline
   \multicolumn{22}{l}{\textit{Panel B: Stocks within industrial sector}} \\
   $1$ && {\color{red} 0.08$^{~~}$} & {\color{red} 0.11$^{~~}$} & {\color{red} 0.06$^{~~}$} & {\color{red} 0.07$^{~~}$} & {\color{red} 0.07$^{~~}$} & {\color{red} 0.04$^{~~}$} & {\color{red} 0.03$^{~~}$} & {\color{red} 0.03$^{~~}$} & {\color{red} 0.03$^{~~}$}&&& {\color{red} 0.10$^{~~}$} & {\color{red} 0.10$^{~~}$} & {\color{red} 0.07$^{~~}$} & {\color{red} 0.08$^{~~}$} & {\color{red} 0.07$^{~~}$} & {\color{red} 0.04$^{~~}$} & {\color{red} 0.04$^{~~}$} & {\color{red} 0.04$^{~~}$} & {\color{red} 0.04$^{~~}$} \\
   $3$ && {\color{red} 0.16$^{~~}$} & {\color{red} 0.09$^{~~}$} & {\color{red} 0.06$^{~~}$} & {\color{red} 0.08$^{~~}$} & {\color{red} 0.07$^{~~}$} & {\color{red} 0.04$^{~~}$} & {\color{red} 0.03$^{~~}$} & {\color{red} 0.03$^{~~}$} & {\color{red} 0.03$^{~~}$}&&& {\color{red} 0.15$^{~~}$} & {\color{red} 0.09$^{~~}$} & {\color{red} 0.07$^{~~}$} & {\color{red} 0.09$^{~~}$} & {\color{red} 0.08$^{~~}$} & {\color{red} 0.04$^{~~}$} & {\color{red} 0.04$^{~~}$} & {\color{red} 0.04$^{~~}$} & {\color{red} 0.03$^{~~}$} \\
   $6$ && {\color{red} 0.01$^{~~}$} & {\color{red} 0.00$^{~~}$} & {\color{red} 0.02$^{~~}$} & {\color{red} 0.04$^{~~}$} & {\color{red} 0.02$^{~~}$} & {\color{blue}-0.01$^{~~}$} & {\color{blue}-0.02$^{~~}$} & {\color{blue}-0.02$^{~~}$} & {\color{blue}-0.02$^{~~}$}&&& {\color{blue}-0.01$^{~~}$} & {\color{red} 0.02$^{~~}$} & {\color{red} 0.05$^{~~}$} & {\color{red} 0.05$^{~~}$} & {\color{red} 0.03$^{~~}$} & {\color{blue}-0.01$^{~~}$} & {\color{blue}-0.01$^{~~}$} & {\color{blue}-0.01$^{~~}$} & {\color{blue}-0.01$^{~~}$} \\
   $9$ && {\color{red} 0.02$^{~~}$} & {\color{red} 0.03$^{~~}$} & {\color{red} 0.05$^{~~}$} & {\color{red} 0.05$^{~~}$} & {\color{red} 0.03$^{~~}$} & {\color{blue}-0.01$^{~~}$} & {\color{blue}-0.01$^{~~}$} & {\color{blue}-0.01$^{~~}$} & {\color{blue}-0.02$^{~~}$}&&& {\color{red} 0.10$^{~~}$} & {\color{red} 0.09$^{~~}$} & {\color{red} 0.08$^{~~}$} & {\color{red} 0.05$^{~~}$} & {\color{red} 0.03$^{~~}$} & {\color{blue}-0.01$^{~~}$} & {\color{blue}-0.01$^{~~}$} & {\color{blue}-0.01$^{~~}$} & {\color{blue}-0.02$^{~~}$} \\
   $12$ && {\color{red} 0.05$^{~~}$} & {\color{red} 0.05$^{~~}$} & {\color{red} 0.05$^{~~}$} & {\color{red} 0.04$^{~~}$} & {\color{red} 0.02$^{~~}$} & {\color{blue}-0.02$^{~~}$} & {\color{blue}-0.02$^{~~}$} & {\color{blue}-0.02$^{~~}$} & {\color{blue}-0.03$^{~~}$}&&& {\color{red} 0.07$^{~~}$} & {\color{red} 0.05$^{~~}$} & {\color{red} 0.02$^{~~}$} & {\color{red} 0.02$^{~~}$} & {\color{red} 0.00$^{~~}$} & {\color{blue}-0.03$^{~~}$} & {\color{blue}-0.03$^{~~}$} & {\color{blue}-0.03$^{~~}$} & {\color{blue}-0.04$^{~~}$} \\
   $24$ && {\color{red} 0.06$^{~~}$} & {\color{red} 0.06$^{~~}$} & {\color{red} 0.04$^{~~}$} & {\color{red} 0.03$^{~~}$} & {\color{red} 0.01$^{~~}$} & {\color{blue}-0.01$^{~~}$} & {\color{blue}-0.00$^{~~}$} & {\color{blue}-0.01$^{~~}$} & {\color{blue}-0.01$^{~~}$}&&& {\color{blue}-0.03$^{~~}$} & {\color{blue}-0.02$^{~~}$} & {\color{blue}-0.02$^{~~}$} & {\color{blue}-0.02$^{~~}$} & {\color{blue}-0.03$^{~~}$} & {\color{blue}-0.03$^{~~}$} & {\color{blue}-0.02$^{~~}$} & {\color{blue}-0.03$^{~~}$} & {\color{blue}-0.03$^{~~}$} \\
   $36$ && {\color{blue}-0.03$^{~~}$} & {\color{blue}-0.02$^{~~}$} & {\color{blue}-0.02$^{~~}$} & {\color{blue}-0.03$^{~~}$} & {\color{blue}-0.04$^{~~}$} & {\color{blue}-0.05$^{~~}$} & {\color{blue}-0.04$^{~~}$} & {\color{blue}-0.05$^{~~}$} & {\color{blue}-0.05$^{~~}$}&&& {\color{blue}-0.04$^{~~}$} & {\color{blue}-0.06$^{~~}$} & {\color{blue}-0.06$^{~~}$} & {\color{blue}-0.06$^{~~}$} & {\color{blue}-0.06$^{~~}$} & {\color{blue}-0.06$^{~~}$} & {\color{blue}-0.07$^{~~}$} & {\color{blue}-0.07$^{~~}$} & {\color{blue}-0.07$^{~~}$} \\
   $48$ && {\color{red} 0.01$^{~~}$} & {\color{blue}-0.01$^{~~}$} & {\color{blue}-0.02$^{~~}$} & {\color{blue}-0.02$^{~~}$} & {\color{blue}-0.04$^{~~}$} & {\color{blue}-0.05$^{~~}$} & {\color{blue}-0.05$^{~~}$} & {\color{blue}-0.05$^{~~}$} & {\color{blue}-0.04$^{~~}$}&&& {\color{blue}-0.06$^{~~}$} & {\color{blue}-0.06$^{~~}$} & {\color{blue}-0.06$^{~~}$} & {\color{blue}-0.07$^{~~}$} & {\color{blue}-0.08$^{~~}$} & {\color{blue}-0.08$^{~~}$} & {\color{blue} \bf -0.08$^{*~}$} & {\color{blue} \bf -0.08$^{*~}$} & {\color{blue} \bf -0.07$^{*~}$} \\
   $60$ && {\color{red} 0.02$^{~~}$} & {\color{red} 0.01$^{~~}$} & {\color{red} 0.00$^{~~}$} & {\color{blue}-0.00$^{~~}$} & {\color{blue}-0.02$^{~~}$} & {\color{blue}-0.03$^{~~}$} & {\color{blue}-0.02$^{~~}$} & {\color{blue}-0.02$^{~~}$} & {\color{blue}-0.01$^{~~}$}&&& {\color{blue}-0.06$^{~~}$} & {\color{blue}-0.05$^{~~}$} & {\color{blue}-0.06$^{~~}$} & {\color{blue}-0.07$^{~~}$} & {\color{blue} \bf -0.09$^{*~}$} & {\color{blue} \bf -0.09$^{*~}$} & {\color{blue} \bf -0.07$^{*~}$} & {\color{blue}-0.07$^{~~}$} & {\color{blue}-0.06$^{~~}$} \\
   \hline
   \end{tabular}
   \label{TB:ChinaIndustrySector:AlterConsume}
\end{table}

Table \ref{TB:ChinaIndustrySector:AlterConsume} shows that, for the industrial sector index of Discretionary Consumption, there are significantly positive excess returns for both the HL and MOP methods when the look-back and holding periods are small, suggesting the presence of the TSMOM effect in the short term. No TSCON effect is observed. In contrast, we observe only the TSCON effect in the long term for individual stocks within the sector, when the MOP method is adopted. Furthermore, the results about the industrial sectors are mixed, which are largely dependent on the different trading signaling methods (HL and MOP) used and on whether the TSMOM strategy are applied to the industrial sector ``index'' or the individual stocks. For certain industrial sectors, no pronounced TSMOM or TSCON effects are observed, which might be partially attributed to our previous finding that the performance of TSMOM (TSCON) strategies is related to firm-specific characteristics.

\begin{figure}[!ht]
\centering
\includegraphics[width=6.5cm]{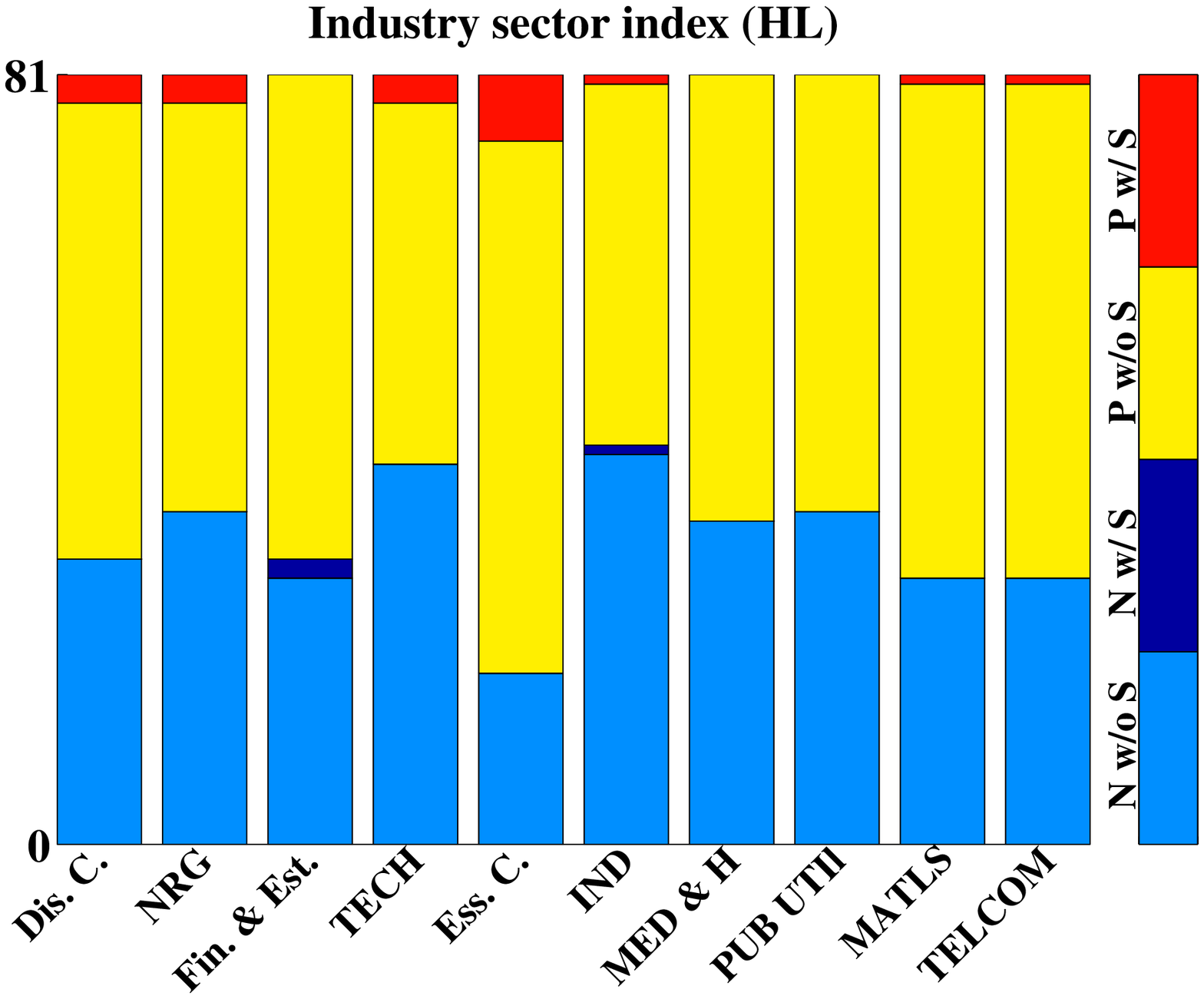}\hspace{3mm}
\includegraphics[width=6.5cm]{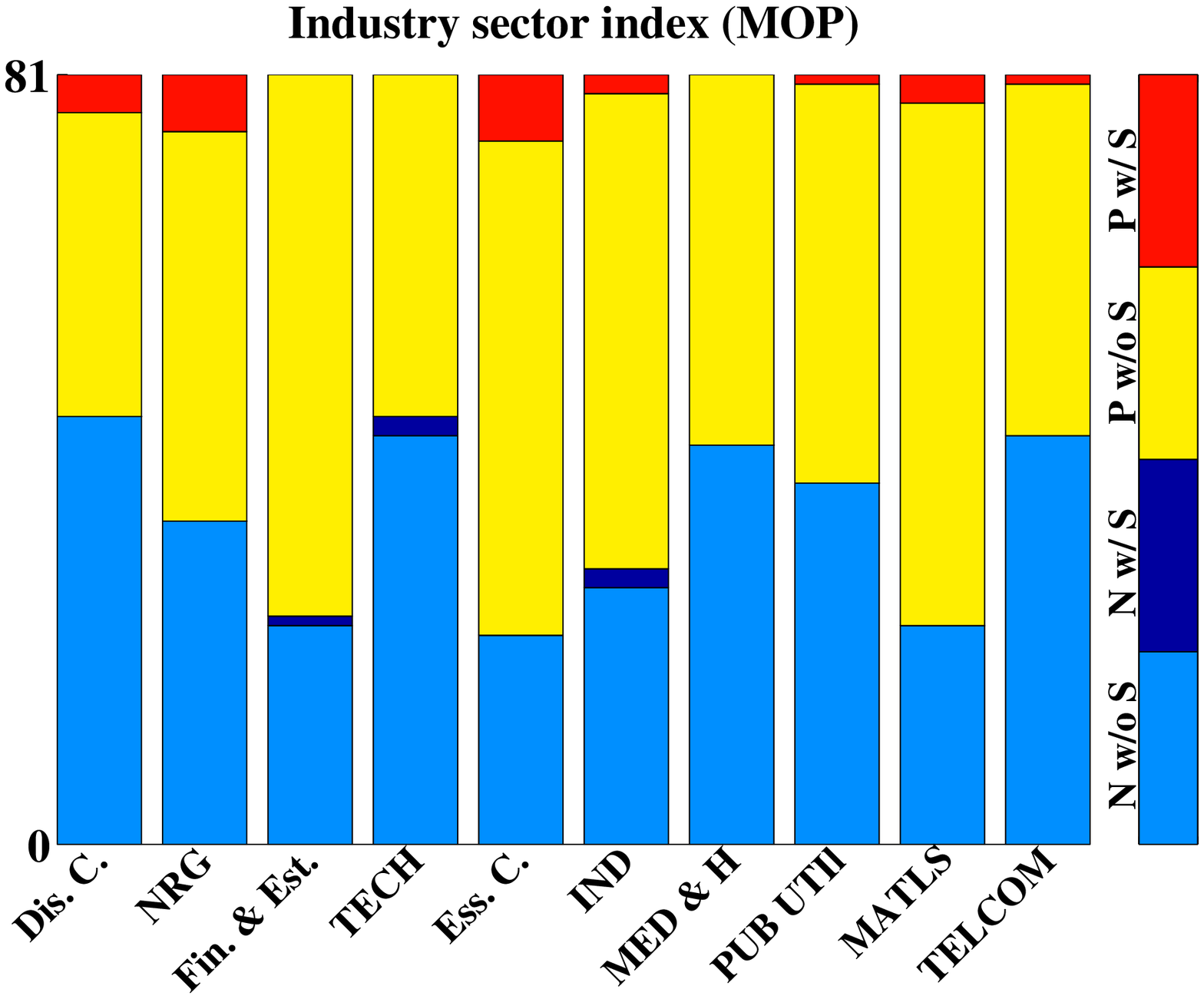}\\
\includegraphics[width=6.5cm]{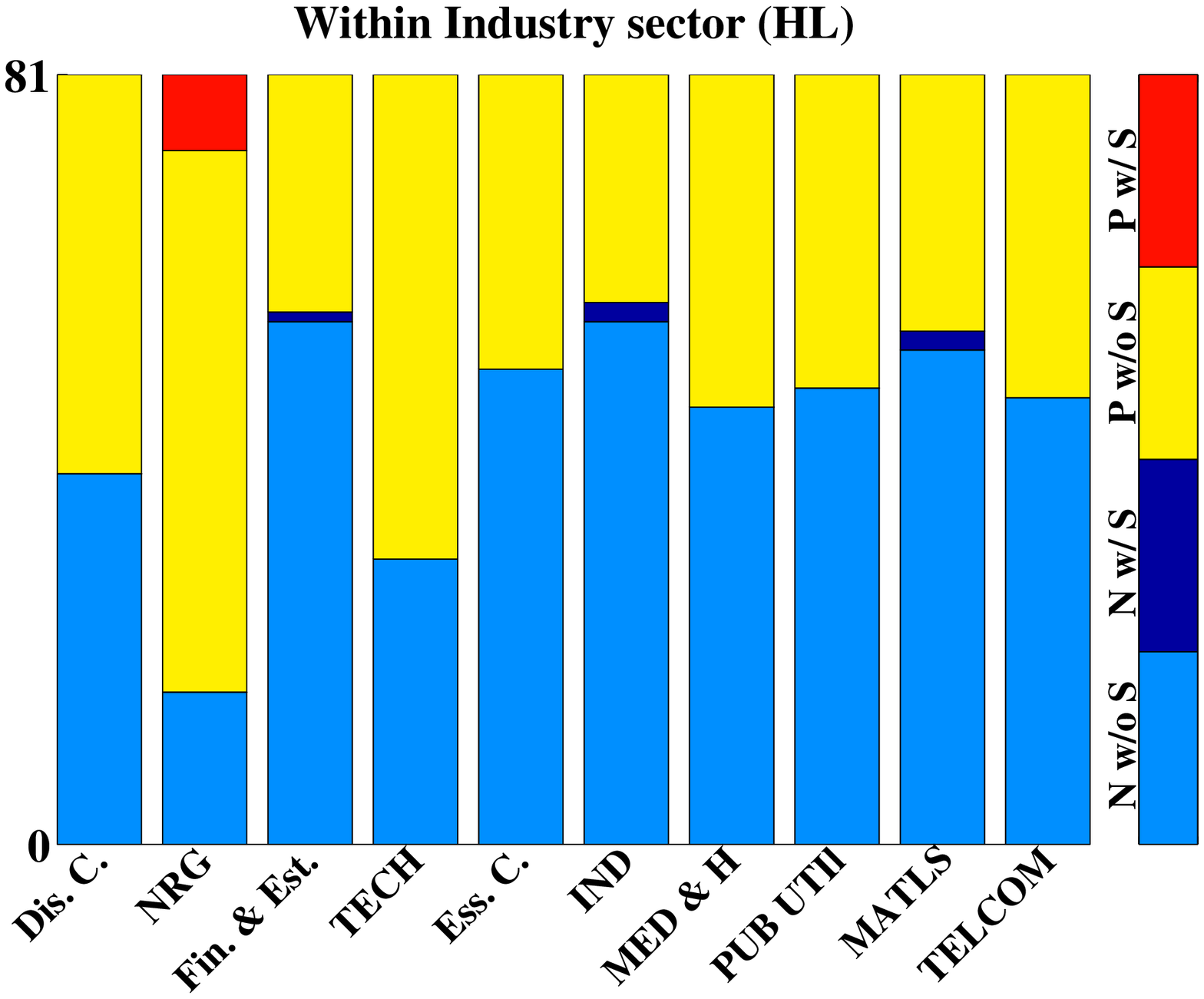}\hspace{3mm}
\includegraphics[width=6.5cm]{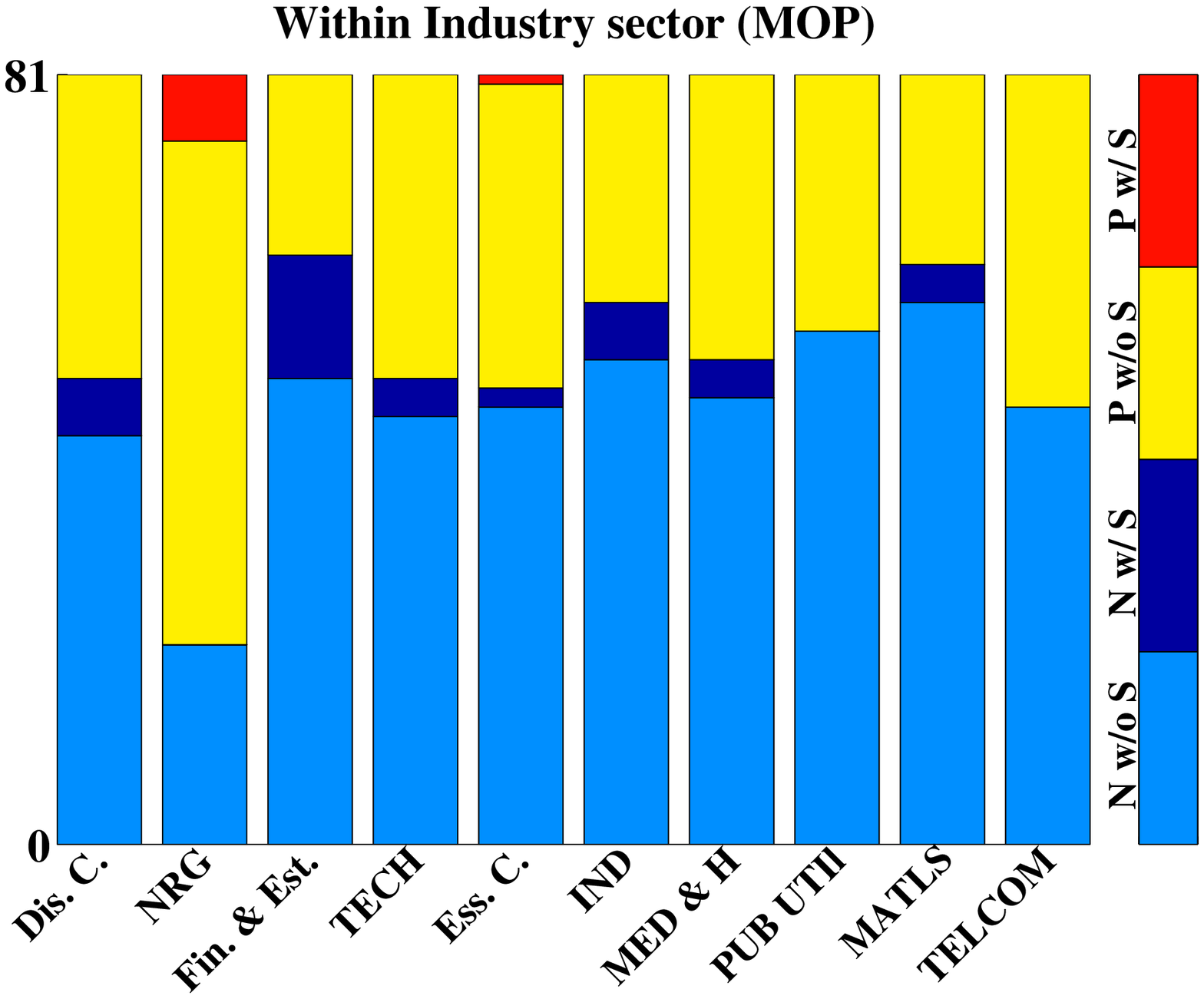}
  \caption{\label{Fig:IndustrySector}(Color online) The respective proportion of significant TSMOM strategies (P w/ S), insignificant TSMOM strategies (P w/o S), significant TSCON strategies (N w/ S), and insignificant TSCON strategies (N w/o S). The top panel is for the results based on industrial sector index and the bottom is for the results based on individual stocks within the industrial sector. The left column corresponds to the results under HL, and the right is for the results under MOP.}
\end{figure}

For further analysis, Fig.~\ref{Fig:IndustrySector} presents the respective proportion of the significant TSMOM strategies, the insignificant TSMOM strategies, the significant TSCON strategies, and the insignificant TSCON strategies. As for the results of industrial sector index in the upper panel of Fig.~\ref{Fig:IndustrySector}, it is evident that the sector of Essential Consumption has the highest number of significant TSMOM strategies, indicating the most pronounced TSMOM effect, which is robust to the trading signaling methods HL and MOP. As for the results based on the individual stocks in the bottom panel of Fig.~\ref{Fig:IndustrySector}, the sector of Energy has the highest number of significant TSMOM strategies, indicating the most pronounced TSMOM effect under HL and MOP. The bottom-right plot shows that several sectors have a significant TSCON effect.

In sum, the results based on the industrial sector index imply a more significant TSMOM effect and the results based on individual stocks imply more pronounced TSCON effects. As is well known, the Chinese stock market is a ``policy-driven'' stock market, where the stock prices are largely influenced by the policies released by the government. Most of the news or policies released are related to some industry and this might lead to the buying (selling) stocks from the same industrial sector with related good (bad) news. Under such circumstance, the stocks in the same industrial sector would move up or fall down in the same direction. Thus, the TSMOM strategy applied to the whole stock group based on the certain industrial sector is more likely to be significantly positive.

\section{Conclusion}
\label{S1:Conclusion}

Our work takes a close look at the time series momentum and contrarian effects in the Chinese stock market.
Firstly, the TSMOM strategies with various look-back and holding periods are applied to three major indexes (SHCI, SZCI and CSI 300) and we observe the different levels of the TSMOM and TSCON effects. Compared with the results in the US market, our findings suggest that the level of market efficiency is higher than the US market, which is seemingly unreasonable. This might be attributed to the data frequency and explained more favorable within the framework of the Adaptive Markets Hypothesis.

We additionally conduct the study on the TSMOM profitability based on different firm-specific characteristics and industrial sectors. We find the profitability of the TSMOM strategies is related to firm-specific characteristics, which is similar to the findings on the cross-sectional momentum effects in China. Specifically, the TSMOM strategies with higher price (including the closing and adjusted price), higher market value and lower trading volume are more profitable and therefore likely to be more statistical significant. In comparison, we find no evidence supporting the relatively pronounced TSMOM effects for certain industrial sectors£¬indicating the insensitivity of the TSMOM effect to industrial sector.

\section*{Acknowledgement}

This work is partially supported by the National Natural Science Foundation of China (71571121 and 71532009).

\end{document}